\documentclass[12pt]{iopart}
\usepackage{iopams}  
\usepackage{graphicx}
\usepackage{hyperref}
\usepackage{cite}
\begin{document}

\title[Weak Gravitational Field Effects On Large-Scale  Optical Interferometric Bell Tests]{Weak Gravitational Field Effects On Large-Scale  Optical Interferometric Bell Tests}

\author{M Rivera-Tapia$^{1,2}$, A Delgado$^{1,2}$ and G Rubilar$^2$}

\address{$^1$ Instituto Milenio de Investigaci\'on en \'Optica, Universidad de Concepci\'on, Concepci\'on, Chile}
\address{$^2$ Departamento de F\'isica, Facultad Ciencias F\'isicas y Matem\'aticas, Universidad de Concepci\'on,  Concepci\'on, Chile}
\ead{mriverat@udec.cl}
\vspace{10pt}
\begin{indented}
\item[]December 2019
\end{indented}

\begin{abstract}
The technological refinement of experimental techniques has recently allowed the generation of two-photon polarization-entangled states at low Earth orbit, which has been subsequently applied to quantum communications. This achievement paves the way to study the interplay between General Relativity and Quantum Mechanics in new setups. Here, we study the generation of two-photon energy-time entangled states via large scale Franson and Hugged interferometric arrays in the presence of a weak gravitational field. We show that for certain configurations of the arrays, an entangled state emerges as a consequence of the gravitational time delay. We also show that the aforementioned arrays generate entanglement and violate the Clauser-Horne-Shymony-Holt inequality under suitable conditions even in the presence of frequency dispersion.
\end{abstract}

\submitto{\CQG}
%\maketitle
\section{Introduction}
Quantum Mechanics and General Relativity are very successful physical theories. They exhibit a remarkably accurate predictive power ranging from the atomic structure to gravitational waves. Furthermore, contemporary technological applications, such as atomic clocks \cite{Ludlow2015} and the Global Positioning System \cite{Kaplan1996}, find their roots in the aforementioned theories.

The interplay between Quantum Mechanics and General Relativity is a different matter: a quantum theory of gravity is still missing. In spite of this, the interaction between quantum systems and gravity in the weak-field limit has been an intensive research subject, both from theoretical and experimental viewpoints. A canonical example of this is the celebrated Colella-Overhauser-Werner experiment \cite{Colella1975}. In this, a neutron beam propagates inside a Mach-Zehnder interferometer whose arms are at different gravitational potentials. The interference pattern displays a phase difference that is produced by the action of the gravitational field on the neutrons. Similar interferometric setups have been proposed to test cosmological effects, such as, for instance, Sagnac interferometry to measure the rotation parameter entering in G\"odel's metric \cite{Delgado2002, Kajari2004}. Other interferometric studies have been carried out to demonstrate interference of a single photon along a ground station and satellites \cite{Vallone2016}, Wheeler's delayed-choice experiment \cite{Vedovato2017}, and to propose an optical test to study the Einstein equivalence principle \cite{Terno2019}. 

Interferometric visibility has also been proposed as an efficient tool to study proper time in the general relativistic context \cite{Zych2011, Zych2012, Pallister20117}. In this case, the interfering particles are endowed with a clock that is implemented by means of an internal degree of freedom or the Shapiro effect. According to the notion of proper time, clocks will evolve conditionally to the propagation path inside the interferometer, that is, the clocks evolve into different nonorthogonal quantum states. Due to the complementarity between interferometric visibility and which-way information available at the clock states, the former will decrease. This gravitationally induced time dilation has also been proposed as a decoherence mechanism for quantum superpositions \cite{Pikovski2015}.

Here, we study the effect of a weak gravitational field on large scale optical interferometric Bell tests. This is motivated by the constant need to push the validity boundary of Quantum Mechanics as well to explore possible applications of entanglement to satellite-based quantum communications \cite{ Pfennigbauer2005, Yin2017, Liao2018, Ren2017, Liao2017, Oi2017, Kerstel2018, Calderato2019, Xu2019, Feng2019}. In particular, we study Franson \cite{Franson1989} and Hugged \cite{Cabello2009} interferometric arrays. Each one of these consists of a source of twin photons, generated by means of spontaneous parametric down-conversion \cite{Friberg1985,Hong1985,Hong1986,Ghosh1987,Ou1988}, and two Mach-Zehnder interferometers. In the Franson array, each photon enters a different interferometer.
In the Hugged array the interferometers are interlinked in more complex geometry and both twin photons can enter into the same interferometer. The output ports of the interferometers are endowed with single-photon detectors. In the Franson and Hugged arrays, the so-called energy-time entanglement arises due to the simultaneous generation of the twin photons, the impossibility of distinguishing among pairs of propagation paths of the same length via coincidence measurements, and a post-selection process. This discriminates between pairs of photons propagating through paths of equal or different length. It has been shown \cite{Aerts1999,Cabello2009} that the Franson array exhibits a loophole, that is, a local hidden variable theory that emulates the quantum mechanical predictions for the array. This loophole, which is independent of the other loopholes such as locality and detection, arises due to the post-selection of photons arriving at different times. Post-selection requires comparing the detection times of the photons, which makes it possible for the interferometer configurations to depend on each other when the photons arrive at different times. The Hugged array was proposed in Ref. \cite{Cabello2009} and realized in \cite{Lima2010, Cuevas2013}, as a solution to the presence of the loophole in the Franson interferometric array. It has been also shown that if local phases inside the Mach-Zehnder interferometers forming the Franson array can be switched rapidly, that is, in a time scale much shorter than the arrival time difference, there is no such a local hidden variable model. Other modifications of the Franson array have been proposed and experimentally implemented to close the post-selection loophole, such as, for instance, time-bin entanglement \cite{Vedovato2018} and polarization and energy-time hyperentanglement \cite{Strekalov1996}. In spite of the loophole, the Franson array has been experimentally applied to long-distance entanglement-based quantum key distribution (QKD) \cite{Ribordy2000}, high dimensional energy-time QKD \cite{Zhong2015}, and, considering a loophole-free implementation, to high dimensional entanglement distribution in noisy environments \cite{Ecker2019}.

We consider that the Franson and Hugged interferometric arrays experience a weak gravitational field. In particular, we assume that the available propagation paths are at different gravitational potentials and that the gradient of the gravitational field is small enough such that it can be treated as a perturbation. In this scenario, we calculate the gravitational time delays affecting the free propagation of photons along the optical paths forming the interferometric arrays. We impose conditions onto the proper length of the propagation paths such that the only source of temporal delay in the arrival of photons to detectors is the gravitational time dilation. A Franson interferometric array formed by balanced and geometrically identical interferometers does not allow to carry out the post-selection process when placed on a gravity equipotential surface. In this case, all optical paths have the same length and, consequently, all pairs of paths are associated with the simultaneous detection of pairs of photons. Thus, the array does not allow to generate a maximally entangled state. However, when the paths are affected by different gravitational potentials, we show that the gravitational time dilatation generates delays for specific pairs of paths and the post-selection process can be carried out. Thereby, the presence of a weak gravitational field makes possible the generation of a maximally entangled state. 
 
We show that the two-photon states generated by the arrays in presence of a weak gravitational field are maximally entangled and calculate the elapsed time two-photon detection probabilities. These become coincidence detection probabilities when the arrays are at a gravity equipotential surface. The elapsed time detection probabilities are given by a cosine function whose argument is the addition of the gravitational phase shift of each twin photon plus controllable local phases. The phase shifts are given by the gravitational time delay times the frequency. Therefore, the elapsed time detection probabilities are functions of the frequency. For this reason, we also consider the case in which the light source at the interferometric arrays exhibits frequency dispersion. Since detectors do not resolve frequency, the elapsed time detection probabilities must be averaged over the spectral distribution of the light source, which typically conveys a decrease of the two-photon interference visibility. In particular, we consider identical Gaussian distributions for the frequency of each photon. In this case, the elapsed time detection probabilities are given by a harmonic oscillation with an exponentially decreasing amplitude. To study the impact of frequency dispersion on the entanglement of the states generated by the Franson and Hugged interferometric arrays in presence of a weak gravitational field, we resort to the Clauser-Horne-Shymony-Holt (CHSH) inequality $|\Sigma|\le2$ \cite{Bell1964, CHSH1969, Freedman1972, Aspect1981, Tittel1998, Weihs1998, Horodecki2009, Hensen2015, Abellan2019, Handsteiner2017, Rauch2018}. Quantum states that violate this inequality can not be described by an ad-hoc classical theory and are, at least, partially entangled. We show that the functional $\Sigma$ is also given by a harmonic oscillation with an exponentially decreasing amplitude. The scale of the exponential function is given by $A_{*}=\sqrt{\ln 4} \left(c^3/g\sqrt{\sigma^{2}_{1} +\sigma^{2}_{2}} \right)$, where $g$ is gravitational acceleration on the Earth surface, $c$ is the speed of light, and $\sigma_1$ and $\sigma_2$ are the frequency width of the Gaussian distribution of each twin photon generated by the light source. Thereby, Franson and Hugged arrays whose Mach-Zehnder interferometers have a proper area $A$ larger than $A_*$ generate two-photon states that do not violate the CHSH inequality. We also show that for certain configurations of Franson or Hugged arrays the harmonic oscillation can be suppressed, in such a way that the functional $\Sigma$ is solely given by a decreasing exponential function.

This article is organized as follows: In Sec.\thinspace\ref{SEC2} we briefly describe Franson and Hugged interferometric arrays and the CHSH inequality. In Sec.\thinspace\ref{SEC3} we calculate the various time delays and the phase shifts introduced by a weak gravitational field in the interferometric arrays. In Sec.\thinspace\ref{SEC4} we describe quantum state of light generated by the arrays in the presence of a weak gravitational field, calculate the two-photon detection probabilities, and obtain the two-photon interference visibility. In Sec.\thinspace\ref{SEC5} we study the violation of the CHSH inequality in the presence of a weak gravitational field for each array. Finally, in Sec.\thinspace\ref{SEC6} we summarize and conclude.

\section{Franson and Hugged Interferometric Arrays} \label{SEC2}

Franson \cite{Franson1989} and Hugged \cite{Cabello2009} interferometric arrays have been proposed as feasible sources of energy-time (or time-bin) entanglement, which have been subsequently employed to experimentally realize the violation of Bell-like inequalities, such as the CHSH inequality.
Let us consider Figures \ref{Franson} and \ref{Hugged} depicting Franson and Hugged interferometric arrays, respectively, and suppose for now that the gravitational potential difference between the arms of each interferometric array vanishes. Both arrays are composed of two Mach-Zehnder interferometers, a light source, and a single photon detector in each of the four available output ports. In the Franson array the Mach-Zehnder interferometer to the left (right) has two paths $\gamma_1$ ($\gamma'_1$) and $\gamma_2$ ($\gamma'_2$) with proper lengths $L_1$ ($L'_1$) and $L_2+2H$ ($L'_2+2H$), respectively. In the Hugged array the Mach-Zehnder interferometer to the left (right) has two paths $\gamma_1$ ($\gamma'_1$) and $\gamma'_2$ ($\gamma_2$) with proper lengths $L_1$ ($L'_1$) and $L'_2+2H$ ($L_2+2H$), respectively. The optical path difference in each Mach-Zehnder interferometer is given by $\Delta L$. The light source generates at a random time two twin photons by means of spontaneous parametric down-conversion. The source operates in heraldic mode, that is, both twin photons are simultaneously created in a separable state. Both photons have the same polarization state and propagate in opposite directions.

In the case of the Franson interferometric array, the light source is between the Mach-Zehnder interferometers. Thereby, the interferometers are physically disconnected and each photon of the pair always enters a different interferometer. The setup in the Hugged interferometric array is more complex. Here, both interferometers are interlinked, that is, they share a segment of optical path. The source is located at this segment. Consequently, both twin photons might enter the same interferometer.

After the generation of the twin photons, each one of them interacts with a beam splitter. This creates a new state for the photon propagating to the left (right) that corresponds to the coherent superposition of the states $|\gamma_1\rangle$ ($|\gamma'_1\rangle$) and $|\gamma_2\rangle$ ($|\gamma'_2\rangle$) that describe the evolution along paths $\gamma_1$ ($\gamma'_1$) and $\gamma_2$ ($\gamma'_2$), respectively. Thereafter, each photon interacts with a second beam splitter that leads to two output ports. In the case of the Franson array these beam splitters are associated to a single photon described by the orthogonal states $|\gamma_1\rangle\pm e^{i\phi_1}|\gamma_2\rangle$ or $|\gamma'_1\rangle\pm e^{i\phi_2}|\gamma'_2\rangle$, where we have included locally generable phases $\phi_1$ and $\phi_2$, for instance, with the help of a piezoelectric device controlling a translational stage on which the beam splitters are placed \cite{Lima2010}. Thereby, the detection of a single photon does not allow to determine the actual path followed by the detected photon. In the case of the Hugged array we obtain the superpositions $|\gamma_1\rangle\pm e^{i\phi_1}|\gamma'_2\rangle$ and $|\gamma'_1\rangle\pm e^{i\phi_2}|\gamma_2\rangle$ for photons propagating to the left or to the right, respectively.

In Figs. \ref{Franson} and \ref{Hugged} depicting Franson and Hugged interferometric arrays we have included delay lines, which are symbolically depicted as fiber loops (as employed in \cite{Cuevas2013}). Each delay line can be a free-space optical delay line, which consists of a geometric array of retroreflector mirrors placed on a servo-controlled translational stage, or a fiber-based optical delay line, which are based on chirped fiber gratings or coiled fiber. The delay lines allow one to control the optical path length, which in turn controls the optical path difference $\Delta L$ between the arms of the Mach-Zehnder interferometers forming the Franson and Hugged arrays. Thereby, it is possible to balance the interferometers. The delay lines are located next to the beam splitters and at the same gravitational potential.

Non-local coincidences, that is, two simultaneous detections in detectors belonging to different interferometers, allow us the generation of an entangled state. Simultaneity requires that the elapsed time between the generation of a photon and its detection must be equal for both twin photons. This happens only if both photons follow paths of the same length, that is, both twin photons propagate through paths $\gamma_1$ and $\gamma'_1$ or paths $\gamma_2$ and $\gamma'_2$. Thus, after the generation of a pair of twin photons we might have coincidences after a time interval $\Delta t_0$, for photons propagated through paths $\gamma_1$ and $\gamma'_1$, or $\Delta t_0+\Delta L/c$, for photons propagated through paths $\gamma_2$ and $\gamma'_2$. 

Since the light source generates a pure state and the interferometric arrays preserve the purity of the quantum state, the state of the twin photons before the output ports is given by
\begin{eqnarray}
\vert \Phi\rangle &=& \frac{1}{2}\left[ \vert{\gamma_1(\Delta t_0)} \rangle + \vert{\gamma_2(\Delta t_0 + \Delta L/c)} \rangle \right] \left[ \vert{\gamma'_1(\Delta t_0)}\rangle + \vert{\gamma'_2(\Delta t_0 + \Delta L/c)}\rangle \right].
\end{eqnarray}
Since in the spontaneous parametric down conversion process the generation time of a pair of photons is uncertain, it is not possible to distinguish between twin photons that propagated through paths $\gamma_1$ and $\gamma'_1$ or paths $\gamma_2$ and $\gamma'_2$. Furthermore, since these two events are mutually exclusive and the interferometric arrays preserve the coherence, the ensemble of photons detected in coincidence is described by the pure state
\begin{eqnarray}
\vert \Psi\rangle &=& \frac{1}{\sqrt{2}}\left[ \vert{\gamma_1(\Delta t_0)}\rangle \vert{\gamma'_1(\Delta t_0)}\rangle  + \vert{\gamma_2(\Delta t_0 + \Delta L/c)}\rangle \vert{\gamma'_2(\Delta t_0 + \Delta L/c)}\rangle \right],
\label{SSLL}
\end{eqnarray}
which is maximally entangled. 

Franson and Hugged interferometric arrays generate the above state by means of a post-selection procedure applied to coincidence detection. Here, indistinguishable events are post-selected, that is, photo-detections occurring at the same time or within a small coincidence window. Photo-detections associated with photons arriving at different times are discarded. In the case of the Franson array, the post-selection requires to compare detection times registered at each Mach-Zehnder interferometer. However, a local hidden variable model admits the local time delays to depend on local parameters, that is, on the locally generable phases $\phi$ and $\phi'$ located at each beam splitter. Thereby, the Franson array is affected by a loophole and measurements carried out on the total ensemble of photons cannot violate the CHSH inequality. In the case of the Hugged array, the post-selection is local, since only photons propagating by similar paths, $(\gamma_1,\gamma'_1)$ paths or $(\gamma_2,\gamma'_2)$ paths, arrive at different interferometers. Consequently, two consecutive detections on the same Mach-Zehnder interferometer indicate a pair of twin photons following a combination of $(\gamma_1,\gamma'_2)$ paths or $(\gamma'_1,\gamma_2)$ paths. Due to this characteristic, the Hugged array allows closing the post-selection loophole exhibited by the Franson array. There are other solutions to close the post-selection loophole: fast switching of local phases \cite{Aerts1999}, the use of time-bin entanglement \cite{Vedovato2018}, and polarization and energy-time hyperentanglement \cite{Strekalov1996}.

The probability of coincidences is given for both interferometric arrays by the expression
\begin{equation}
p_{i,j}(\alpha,\beta) = \frac{1}{4}\left(1 -(-1)^{\delta_{ij}}V \cos(\alpha + \beta) \right),
\label{ProbFRHG-NORMAL}
\end{equation}
where $i,j=\pm 1$ label the output ports of the Mach-Zehnder interferometers ($i$ at the left and $j$ at the right of the SPDC source), $\alpha$ and $\beta$ are the local phase shifts, and $V$ is the two-photon visibility (that is assumed equal for all pairs of detectors).  

The state generated by Franson and Hugged interferometric arrays has been employed to study the non-local nature of Quantum Mechanics via the test of the CHSH inequality \cite{CHSH1969}. This is given by the expression
\begin{equation}
|\Sigma|\leq 2,
\label{CHSH}
\end{equation}
where the functional $\Sigma$ is
\begin{equation}
\Sigma = E(\alpha,\beta) + E(\alpha',\beta)+E(\alpha,\beta')-E(\alpha',\beta')
\label{CHSH1}
\end{equation}
and
\begin{eqnarray}
E(\alpha,\beta) = p_{+,+}(\alpha,\beta) + p_{-,-}(\alpha,\beta) - p_{+,-}(\alpha,\beta)  - p_{-,+}(\alpha,\beta)
\label{DICHOTOMIC}
\end{eqnarray}
is the expectation value of a dichotomic observable with eigenvalues $\pm 1$ that depends on the local phases $\alpha$ and $\beta$. If the probabilities have the form in Eq. (\ref{ProbFRHG-NORMAL}) above, then the expectation value (\ref{DICHOTOMIC}) becomes $E(\alpha, \beta) = V \cos(\alpha + \beta)$. If the values of the local phases are $\alpha=-\alpha'=\pi/4$, $\beta=0$, and $\beta'=\pi/2$ then the value of $\Sigma$ reduces to $\Sigma=2\sqrt{2}V$. When the visibility is maximal and the generated estate is maximally entangled we obtain $|\Sigma|=2\sqrt{2}$

\section{Gravitational Phase shift in Franson and Hugged Interferometric Arrays}
\label{SEC3}

In this section, we study the propagation of photons inside the Franson and Hugged interferometric arrays considering the action of a weak gravitational field. For this purpose, we review the use of photons as clocks and the determination of the corresponding gravitational time delay and phase shift. Afterward, we apply this to the Franson and Hugged interferometric arrays.

\subsection{Photons as clocks}

In this subsection, we describe a photon as a clock and the effects of the gravitational potential on light. We describe the gravitational field using the Schwarzschild metric in isotropic coordinates since this ensures that our expressions remain unchanged, up to first order in the distances and/or coordinates differences if the interferometric arrays are rotated.

The Schwarzschild metric in isotropic coordinates \cite{Brodutch2015} $(ct,x,y,z)$  is given, to first order in $\phi/c^2$, by the expression
\begin{eqnarray}
ds^2 &=& \left( 1+ 2\frac{\phi(r)}{c^2} \right) c^2 dt^2 - \left(1-2\frac{\phi(r)}{c^2} \right) \left( dx^2 + dy^2 + dz^2 \right),
\label{Metric1}
\end{eqnarray}
where $\phi(r)$ is the Newtonian gravitational potential, given by $\phi(z) = gz+ const$. Since photons propagate along a light-like curve from position $x_0$ to position $x_f$, the corresponding interval of coordinate time is given by
\begin{eqnarray}
\Delta t &=& \pm \left(\frac{1}{c}\right) \int_{x_0}^{x_f} \left(1-2\frac{\phi(r)}{c^2} \right) dl, \label{photonTempCoord}
\end{eqnarray}
where $dl=\sqrt{dx^2 + dy^2 + dz^2}$.  According to Eq.\thinspace(\ref{photonTempCoord}), the temporal coordinate of photons depends on the gravitational potential. Consequently, photons moving through paths with the same change of spatial coordinates under the action of different gravitational potentials have different coordinate time intervals.

Another effect of the gravitational field on a photon is a phase shift. The phase shift of the electromagnetic field describing the photon can be computed within the geometrical optics approximation \cite{Will2010}. In this context, the phase of the electromagnetic potential satisfies the eikonal equation, which corresponds to the Hamilton-Jacobi equation for massless particles on a particular background metric. For the case of a stationary spacetime, the phase shift is given by $\Delta \varphi = \omega_{\infty} \Delta t$ \cite{Brodutch2015, Brodutch2011}, where $\omega_{\infty}$  is the frequency of the electromagnetic field as measured by an observer at infinity, where the gravitational field vanishes, and $\Delta t$ is the corresponding coordinate time interval spent by light on its path. For an observer at rest at some particular location, we can write $\Delta \varphi = \omega \Delta \tau$, where $\omega=\omega_{\infty}/\sqrt{g_{00}}$ is the frequency of the electromagnetic field measured at the position of the observer/clock, and $\Delta \tau = \sqrt{g_{00}}\Delta t$ is the corresponding proper time.
 
The phase shift can be observed using a single photon propagating in a Mach-Zehnder interferometer \cite{Zych2012} whose arms experience different gravitational potentials. This leads, under the approximations described above and in the weak gravitational field limit, to a phase shift $\Delta \varphi = \omega \Delta \tau \approx g\omega H\Delta L/c^3$ in the interference pattern, where $\Delta L$ and $H$ are the characteristic proper length and proper height of the interferometer. Besides the phase shift, a second interesting effect arises: the visibility of the interference pattern decreases exponentially with the square of the difference $\Delta \tau$ of proper time. In the Newtonian limit, the difference of proper time vanishes. Consequently, the visibility is maximal and only the phase shift is present. However, in the context of General Relativity in the weak field limit the difference of proper time is in general non-null and both effects, phase shift and decrease of visibility, are present. 
\subsection{Franson interferometric array}
\begin{figure}
    \centering
        \includegraphics[width=0.8\textwidth]{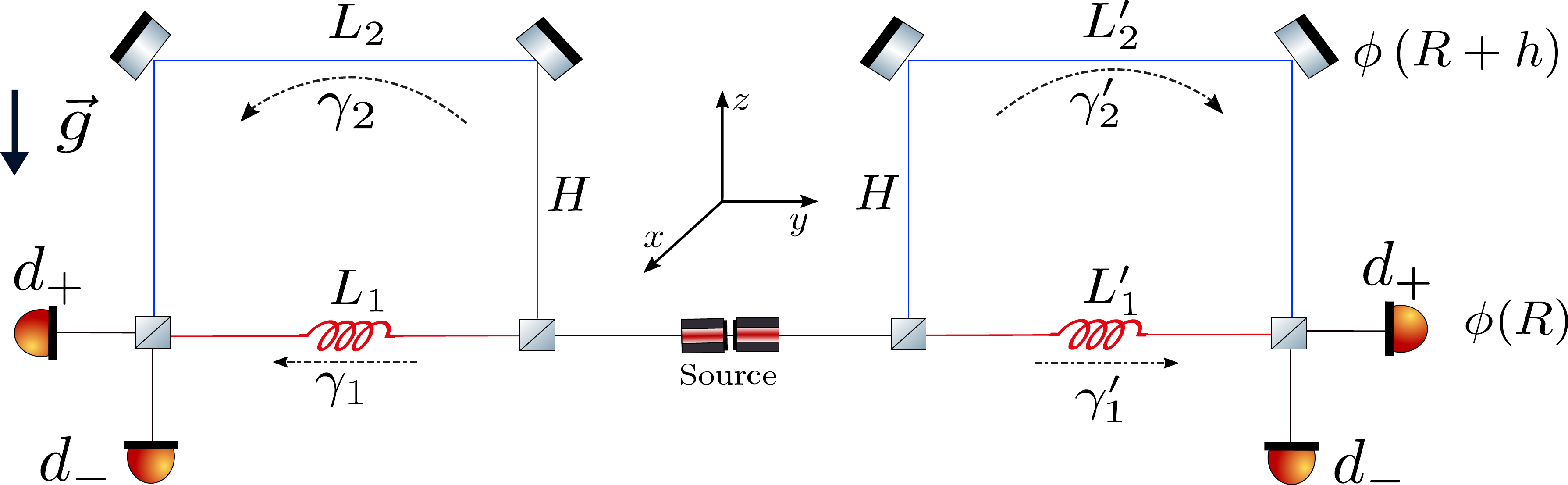} % first figure itself     
          \caption{ Franson interferometric array. A light source concatenates two Mach-Zehnder interferometers. Light source paths $\gamma_1$ and $\gamma'_1$ are located at a gravitational potential $\phi(R)$. The horizontal segments of paths $\gamma_2$ and $\gamma'_2$ are placed at a gravitational potential $\phi(R+h)$. $L_1$ and $L'_1$ indicate the proper length of the horizontal paths $\gamma_1$ and $\gamma'_1$, respectively. $L_2$ and $L'_2$ indicate the proper length of the horizontal segment of paths $\gamma_2$ and $\gamma'_2$, respectively. $H$ denotes the proper length of the vertical segments of paths $\gamma_2$ and $\gamma'_2$. Paths $\gamma_1$ and $\gamma'_1$ contain symbolic delay lines to control de difference of proper length between  paths $\gamma_1$ and $\gamma_2$ and between paths $\gamma'_1$ and $\gamma'_2$. Delay lines are at gravitational potential $\phi(R)$. Detectors  at the left and right output ports of the interferometers are indicated by $d_{+}$ and $d_{-}$.}
          \label{Franson}
\end{figure}

In the Franson interferometric array, the paths $\gamma_1$ and $\gamma'_1$ experience a gravitational potential $\phi(R)$ and the horizontal segments of the paths $\gamma_2$ and $\gamma'_2$  experience a gravitational potential $\phi(R+h)$, where $h=\Delta z$ is the difference in the $z$ coordinate (along the vertical direction) of the upper and lower paths of the interferometer. The vertical segments of paths $\gamma_2$ and $\gamma'_2$ experience a continuous variation of gravitational potential from $\phi(R)$ to $\phi(R +h)$. 

According to the Schwarzschild metric Eq.\thinspace(\ref{Metric1}), the proper length of path $\gamma_1$ is approximately given by
\begin{eqnarray}
L_{1} & \approx & \left( 1- \frac{\phi(R)}{c^2} \right)\Delta x_{_1},
\end{eqnarray}
where $\Delta x_{_1}$ is the interval of spatial coordinates of $\gamma_1$. Similarly, for the horizontal segment of path $\gamma_2$ the proper length is 
\begin{eqnarray}
L_{2} & \approx &  \left( 1 - \frac{\phi(R+h)}{c^2} \right)\Delta x_{_2},
\end{eqnarray}
where $\Delta x_{_2}$ is the interval of spatial coordinates of $\gamma_2$.
The vertical segments of path $\gamma_2$ have a proper height given by
\begin{eqnarray}
H & \approx & \left( 1 -\frac{\phi(R)}{c^2} \right)h. \label{ProperHFranson1}
\end{eqnarray}  
The coordinate time interval Eq.\thinspace(\ref{photonTempCoord}) for a vertical section of the path $\gamma_2$ is, up to first order in $h$, given by
\begin{eqnarray}
\Delta t^{V}_{\gamma_2} & \approx & \frac{h}{c}\left( 1 -2 \frac{\phi(R)}{c^2}\right),
\end{eqnarray}
while the coordinate time intervals for the path $\gamma_1$ and the horizontal segment of path $\gamma_2$ are given by
\begin{equation}
\Delta t_{\gamma_1} \approx \frac{L_{1}}{c}\left(1- \frac{\phi(R)}{c^2} \right) 
\end{equation}
and
\begin{equation}
\Delta t^{H}_{\gamma_2} \approx  \frac{L_{2}}{c}\left(1-\frac{\phi(R+h)}{c^2}\right),
\end{equation} 
respectively.  The total coordinate time interval $\Delta t_{\gamma_2}$ along path $\gamma_2$ is $\Delta t_{\gamma_2} = \Delta t^{H}_{\gamma_2} + 2 \Delta t^{V}_{\gamma_2}$. The elapsed time between generation and detection of a photon that propagates along path $\gamma_1$ or path $\gamma_2$, measured by a clock at the position of the detectors at a gravitational potential $\phi(R)$, is given by the proper time intervals
\begin{equation}
\Delta \tau_{\gamma_1}=\sqrt{1+\frac{2\phi(R)}{c^2}}\Delta t_{\gamma_1} \approx  \frac{L_{1}}{c} 
\label{taugamma1}
\end{equation}
and
\begin{equation}
\Delta \tau_{\gamma_2}=\sqrt{1+\frac{2\phi(R)}{c^2}}\Delta t_{\gamma_2} \approx  \frac{L_{2}}{c}\left(1-\frac{gH}{c^2} \right) + 2\frac{H}{c}, 
\label{taugamma2}
\end{equation}
respectively. Here we used $\phi(R+h)\approx\phi(R)+gh$, to first order in $h$. Analogously, for photons propagating along path $\gamma'_1$ or path $\gamma'_2$ we obtain
\begin{equation}
\Delta \tau_{\gamma'_1}=\sqrt{1+\frac{2\phi(R)}{c^2}}\Delta t_{\gamma'_1} \approx  \frac{L'_1}{c} 
\label{taugamma1'}
\end{equation}
and
\begin{equation}
\Delta \tau_{\gamma'_2}=\sqrt{1+\frac{2\phi(R)}{c^2}}\Delta t_{\gamma'_2} \approx  \frac{L'_2}{c}\left(1-\frac{g H}{c^2} \right) + 2\frac{H}{c}, 
\label{taugamma2'}
\end{equation}
correspondingly.

The differences of proper time $\Delta\tau_{\gamma_1\gamma'_1}=\Delta \tau_{\gamma_1}-\Delta \tau_{\gamma'_1}$ and $\Delta\tau_{\gamma_2\gamma'_2}=\Delta \tau_{\gamma_2}-\Delta \tau_{\gamma'_2}$ describe the elapsed time between the successive detection of two twin photons propagating along paths $(\gamma_1,\gamma'_1)$ and $(\gamma_2,\gamma'_2)$, respectively. Analogously, $\Delta\tau_{\gamma_1\gamma_2}=\Delta \tau_{\gamma_1}-\Delta \tau_{\gamma_2}$ and $\Delta\tau_{\gamma'_1\gamma'_2}=\Delta \tau_{\gamma'_1}-\Delta \tau_{\gamma'_2}$ describe the difference of the time of flight of a single photon that propagates along the arms of each Mach-Zehnder interferometer. The time delays $\Delta\tau_{\gamma_1\gamma'_1}$, $\Delta\tau_{\gamma_2\gamma'_2}$, $\Delta\tau_{\gamma_1\gamma_2}$ and $\Delta\tau_{\gamma'_1\gamma'_2}$ obey the following constraint
\begin{equation}
\Delta\tau_{\gamma_1\gamma'_1}-\Delta\tau_{\gamma_2\gamma'_2}=\Delta\tau_{\gamma_1\gamma_2}-\Delta\tau_{\gamma'_1\gamma'_2}.
\label{Constraint}
\end{equation}

We impose the following condition 
\begin{equation}
\Delta\tau_{\gamma_1\gamma'_1}=\Delta\tau_{\gamma_2\gamma'_2}=\Delta\tau,
\label{Condition}
\end{equation}
with $\Delta\tau\geq0$. This guarantees that it is not possible to distinguish a pair of photons propagating along paths $(\gamma_1, \gamma'_1)$ from a pair of photons propagating along paths $(\gamma_2, \gamma'_2)$ by means of the elapsed time between successive detections. Thereby, the Franson interferometric array generates a maximally entangled state even when there are no coincidences between pairs of detectors at different Mach-Zehnder interferometers. 

According to the constraint Eq.\thinspace(\ref{Constraint}) and the condition Eq.\thinspace(\ref{Condition}), the equality $\Delta\tau_{\gamma_1\gamma_2}=\Delta\tau_{\gamma'_1\gamma'_2}$ must hold. This also constraints the proper lengths. We obtain
\begin{eqnarray}
L_{1}&\approx&L'_1+c\Delta\tau, \label{L1L1p}
\\
L_{2}&\approx&L'_2+c\Delta\tau(1+\frac{gH}{c^2}). \label{L2L2p}
\end{eqnarray} 
Using these expressions, the temporal delays $\Delta\tau_{\gamma_1\gamma_2}$ and $\Delta\tau_{\gamma'_1\gamma'_2}$ become
\begin{eqnarray}
\Delta\tau_{\gamma_1\gamma_2}&=&\frac{L'_1}{c}-\frac{L'_2}{c}(1-\frac{gH}{c^2})-\frac{2H}{c},\\
\Delta\tau_{\gamma'_1\gamma'_2}&=&\Delta\tau_{\gamma_1\gamma_2}.
\end{eqnarray}
Now, we can remove the terms that do not involve the gravitational field by demanding that
\begin{equation}
L'_1=L'_2+2H,
\label{NOGEO}
\end{equation}
which indicates that the proper lengths for paths $\gamma'_1$ and $\gamma'_2$ are equal, that is, the Mach-Zehnder interferometer for paths $\gamma'_1$ and $\gamma'_2$ is balanced. Notice that this means, if one would implement this array using optical fibers to guide the photos, that one would require two fibers, each of length $L'_1$ given by (\ref{NOGEO}), which would then be arranged to form the paths $\gamma'_1$ and $\gamma'_2$; the first one always laying in the horizontal plane, while the second one forming the vertical rectangle ilustrated in Fig. \ref{Franson}.  Under condition Eq.\thinspace(\ref{NOGEO}), the time delays $\Delta\tau_{\gamma_1\gamma_2}$ and $\Delta\tau_{\gamma'_1\gamma'_2}$ reduce to an expression proportional to the gravitational field \cite{Zych2012,Chen2018}, that is, the gravitational time delays are given by
\begin{equation}
\Delta \tau_{\gamma_1,\gamma_2}=\Delta\tau_{\gamma'_1\gamma'_2}\approx \frac{L'_2 g H}{c^3}.
\label{deltatauF}
\end{equation}
Consequently, the phase shift becomes
\begin{eqnarray}
\Delta \varphi &=& \omega \Delta \tau_{\gamma_1\gamma_2} = \omega \Delta \tau_{\gamma'_1\gamma'_2}  \approx  \frac{\omega L'_2 g H}{c^3}.\label{ProperTimeFranson1}
\end{eqnarray}

In order to generate a maximally entangled state, we must be able to perform the post-selection procedure, that is, to distinguish pairs of twin photons following paths $(\gamma_1,\gamma'_1)$ or $(\gamma_2,\gamma'_2)$ from pairs of twin photons following paths $(\gamma_1,\gamma'_2)$ or $(\gamma_2, \gamma'_1)$. Thus, we need to obtain the value of the time delays $\Delta\tau_{\gamma_1\gamma'_2}=\Delta\tau_{\gamma_1}-\Delta\tau_{\gamma'_2}$ and $\Delta\tau_{\gamma_2\gamma'_1}=\Delta\tau_{\gamma_2}-\Delta\tau_{\gamma'_1}$, which can be cast as
\begin{eqnarray}
\Delta\tau_{\gamma_1\gamma'_2}&=&\Delta\tau_{\gamma_1\gamma'_1}+\Delta\tau_{\gamma'_1\gamma'_2},
\\ 
\Delta\tau_{\gamma_2\gamma'_1}&=&\Delta\tau_{\gamma_2\gamma'_2}-\Delta\tau_{\gamma'_1\gamma'_2}.
\end{eqnarray}
Using Eqs.\thinspace(\ref{Condition}) and (\ref{deltatauF}) we can write
\begin{eqnarray}
\Delta\tau_{\gamma_1\gamma'_2} &\approx& \Delta\tau+\frac{L'_2 g H}{c^3},
\\
\Delta\tau_{\gamma_2\gamma'_1} &\approx& \Delta\tau-\frac{L'_2 g H}{c^3},
\end{eqnarray}
which indicate that twin photons following paths $(\gamma_1,\gamma'_2)$ can be distinguished from twin photons following paths $(\gamma_2,\gamma'_1)$ by measuring the elapsed time between successive detections.

%Since twin photons propagating along paths $(\gamma_1,\gamma'_1)$ or $(\gamma_2,\gamma'_2)$ experience a time delay $\Delta\tau$, these can be distinguished from twin photons propagating along paths $(\gamma_1,\gamma'_2)$ or $(\gamma_2, \gamma'_1)$ by measuring the elapsed time between detections. Furthermore, it is the gravitational time delay $L'_2 g H/c^3$ that makes the post-selection process possible and leads to the generation of an entangled state. 

\subsection{Hugged interferometric array}

\begin{figure}
		\centering
        \includegraphics[width=0.7\textwidth]{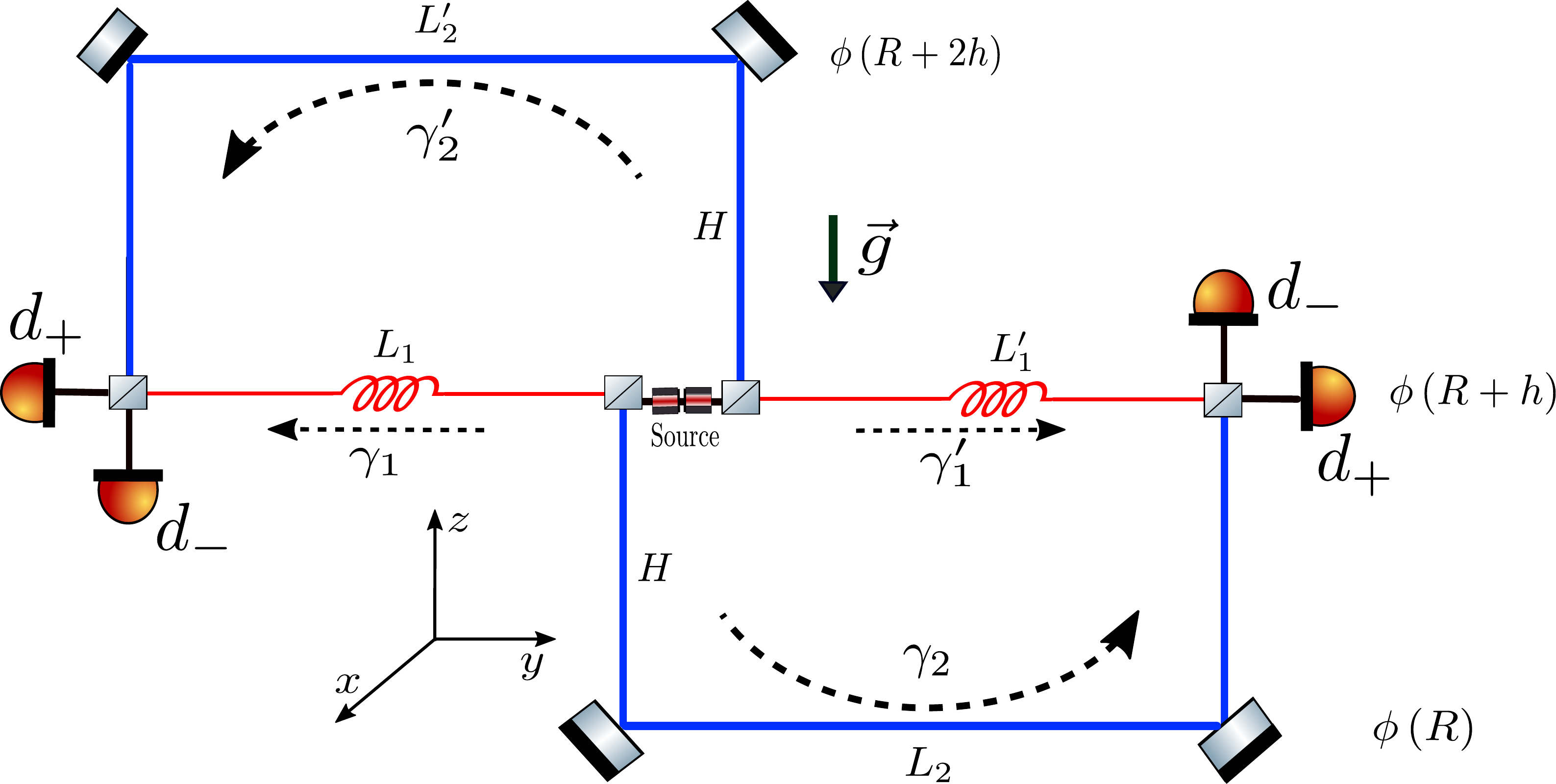} % second figure itself       
         \caption{Hugged interferometric array. A light source is placed on a segment belonging to two Mach-Zehnder interferometers. Light source and paths $\gamma_1$ and $\gamma'_1$ are located at a gravitational potential $\phi(R+h)$. The horizontal segments of paths $\gamma_2$ and $\gamma'_2$ are placed at a gravitational potential $\phi(R+2h)$ and $\phi(R)$, respectively. $L_1$ and $L'_1$ indicate the proper length of the horizontal paths $\gamma_1$ and $\gamma'_1$, respectively. $L_2$ and $L'_2$ indicate the proper length of the horizontal segment of paths $\gamma_2$ and $\gamma'_2$, respectively. $H$ denotes the proper length of the vertical segments of paths $\gamma_2$ and $\gamma'_2$. Paths $\gamma_1$ and $\gamma'_1$ contain symbolic delay lines to control de difference of proper length between  paths $\gamma_1$ and $\gamma_2$ and between paths $\gamma'_1$ and $\gamma'_2$. Delay lines are at gravitational potential $\phi(R+h)$. Detectors  at the left and right output ports of the interferometers are indicated by $d_{+}$ and $d_{-}$. }
         \label{Hugged}
\end{figure}

The Hugged interferometric array, depicted in Fig.\thinspace\ref{Hugged}, has its light source placed at a gravitational potential $\phi(R+h)$. Photons travelling along paths $\gamma_1$ and $\gamma'_1$ also experience this potential. Photons traveling along the horizontal segments of the paths $\gamma'_2$ and $\gamma_2$ experience gravitational potentials $\phi(R+2h)$ and $\phi(R)$, respectively. Vertical segments of path $\gamma'_2$ experience a continuous variation of gravitational potential from $\phi(R+h)$ to $\phi(R +2h)$. Vertical segments of path $\gamma_2$ experience a continuous variation of gravitational potential from $\phi(R)$ to $\phi(R +h)$. 

As in the case of the Franson array, to generate a two-photon entangled state, we impose the condition Eq.\thinspace(\ref{Condition}), which guarantees the indistinguishability of twin photons travelling along paths $(\gamma_1,\gamma'_1)$ from those travelling along paths $(\gamma_2,\gamma'_2)$. 

We can now impose the condition Eq.\thinspace(\ref{NOGEO}), which removes the geometric terms entering in $\Delta\tau_{\gamma_1\gamma_2}$. Thereby, we obtain (for further details, see \ref{A1})
\begin{equation}
\Delta \tau_{\gamma_1\gamma_2}=\Delta \tau_{\gamma'_1\gamma'_2}\approx \frac{L'_2 g H}{c^3}.
\end{equation}
The phase shift becomes
\begin{eqnarray}
\Delta \varphi &=& \omega \Delta \tau_{\gamma_1\gamma_2} = \omega \Delta \tau_{\gamma'_1\gamma'_2}  \approx  \frac{\omega L'_2 g H}{c^3}.
\end{eqnarray}

\section{Two-photon State and Detection Probabilities in Franson and Hugged Interferometric arrays in a weak Gravitational Field}
\label{SEC4}

In the previous section, we calculated various time delays between detection events in the Franson and Hugged interferometric arrays in the presence of a weak gravitational field. In this section, we calculate the effect of temporal delays on the probability of detection between pairs of detectors.

The light source simultaneously generates two photons that propagate in opposite directions. We  assume that the two-photon state generated by the light source is described by the superposition
\begin{equation}
\vert{\psi}\rangle = \int \int d\omega_1 d\omega_2 f(\omega_1,\omega_2) a^{\dagger}(\omega_1) a^{\dagger}(\omega_2) \vert{0}\rangle,
\label{InitialStateFRHG}
\end{equation}
where $\vert 0\rangle$ is the vacuum state of the electromagnetic field and the operator $a^{\dagger}(\omega_i)$ creates a photon that propagates to the left ($i=1$) or to the right ($i=2$) of the light source with frequency $\omega_i$. The spectral function $f(\omega_1,\omega_2)$ depends on the specific spectral properties of the light source and satisfies the normalization condition 
$\int \int d\omega_1d\omega_2\vert f(\omega_1,\omega_2)\vert^2=1$.

The state after a post-selection procedure can be cast as the following coherent superposition (for further details, see \ref{A2}). 
\begin{equation}
|\Psi\rangle=\int \int d\omega_1 d\omega_2 \tilde f(\omega_1, \omega_2)|\psi\rangle,
\end{equation}
with the probability amplitude distribution $\tilde f(\omega_1, \omega_2)$ given by
\begin{equation}
\tilde f(\omega_1, \omega_2)=\frac{1}{\sqrt{2}}f(\omega_1, \omega_2)e^{-i\omega_1\Delta\tau}e^{i(\omega_1+\omega_2)\Delta\tau_{\gamma'_1}},
\end{equation}
and the two-photon state $|\psi\rangle$ defined as
\begin{eqnarray}
|\psi\rangle&=&\frac{1}{\sqrt{2}}(|\omega_1,\gamma_1\rangle|\omega_2,\gamma'_1\rangle
 +e^{i(\omega_1+\omega_2)\Delta\tau_{\gamma'_2\gamma'_1}+i(\alpha +\beta)}
|\omega_1,\gamma_2\rangle|\omega_2,\gamma'_2\rangle),
\label{MAX-STATE}
\end{eqnarray}
where the states $|\omega_1,\gamma_i\rangle|\omega_2,\gamma'_i\rangle$ describe a single photon of frequency $\omega_1$ propagating in path $\gamma_i$ and a single photon of frequency $\omega_2$ propagating in path $\gamma'_i$. 

Let us consider a light source free of frequency dispersion, that is, $f(\omega_1,\omega_2)=\delta(\omega_1-\bar\omega_1)\delta(\omega_2-\bar\omega_2)$, and the case $\alpha=\beta=0$. When all propagation paths are at the same gravitational potential, we can consider that, by a suitable redefinition of the gravitational potential, $\Delta\tau_{\gamma'_1\gamma'_2}$ vanishes. In this case, the interferometric array generates the maximally entangled state
\begin{equation}
\frac{1}{\sqrt{2}}(|\bar\omega_1,\gamma_1\rangle|\bar\omega_2,\gamma'_1\rangle+|\bar\omega_1,\gamma_2\rangle|\bar\omega_2,\gamma'_2\rangle),
\end{equation}
provided that the post-selection process can be carried out. This is the well-known state generated by the Franson and Hugged interferometric arrays. The presence of a weak gravitational field leads to a relative phase in state $|\psi\rangle$, which does not change the amount of entanglement. Thus, the state $|\Psi\rangle$ is a coherent superposition of maximally entangled two-photon states.

An interesting effect arises when considering a Franson interferometric array in a gravity equipotential surface. This is equivalent to consider $g=0$ in the expressions above. Moreover, we consider the case in which the two Mach-Zehnder interferometers are balanced, which means $\Delta\tau=0$ in Eqs.\ (\ref{L1L1p}) and (\ref{L2L2p}). In this case, the four possible combinations of paths have exactly the same length and consequently it is not possible to distinguish among them. Therefore, the array only generates a separable state. We can, however, rotate the array in such a way that the arms experience different gravitational potentials but the Mach-Zehnder interferometers stay balanced and geometrically equivalent with respect to the proper lengths. This is ensured by our choice of the Schwarzschild metric in isotropic coordinates. 
In this case we have that $\Delta\tau$ vanishes and the gravitational time delays become $\Delta\tau_{\gamma_1\gamma'_2}=L'_2 g H/c^3=-\Delta\tau_{\gamma_2\gamma'_1}$. Thereby, the gravitational time delays do not vanish, the post-selection process is still possible, and the array generates a maximally entangled state. In this particular case, the presence of a weak gravitational field makes possible the generation of a maximally entangled state.

In the case of the Hugged interferometric array, if we consider a balanced and geometrically equivalent array on an equipotential surface, the arrival times to each detector are identical for all paths in the array. However,  it is still possible to generate an entangled state. This is due to the fact that the post-selection procedure is local. Since the two paths in each Mach-Zehnder interferometer have equal length both photons coalesce and are detected at the same detector. In this case, we can discard this class of events. Coincidence detections at different Mach-Zehnder interferometers are described by a maximally entangled. This corresponds to the superposition of two orthogonal states, each one describing macroscopically distinguishable propagation paths.

If we consider a 90º rotation in the vertical of the Hugged interferometric array that is formed by balanced and geometrically equivalent Mach-Zehnder interferometers, as described in previous paragraph, we obtain that the proper lengths satisfy the following equations
\begin{eqnarray}
L_{1} & \approx & L'_1, \\
L_{2} & \approx & \left(1 + 2gH/c^2 \right)L'_2.
\end{eqnarray}
Then, for the temporal delay along each path we have that
\begin{eqnarray}
\Delta \tau_{\gamma_1} - \Delta \tau_{\gamma'_1} &=& 0, \\
\Delta \tau_{\gamma_2} - \Delta \tau_{\gamma'_2} & = & 4 g L'_2H/c^3.
\end{eqnarray}
Thereby, photons propagating through the paths $(\gamma_1,\gamma'_1)$ are detected in coincidence. Photons propagating through the paths $(\gamma_2,\gamma'_2)$ are detected with a time delay. In this case we can distinguish between photons propagating through paths $(\gamma_1,\gamma'_1)$ or paths $(\gamma_2,\gamma'_2)$. In absence of frequency dispersion, according Eq. (\ref{MAX-STATE}), the array still generates a maximally entangled state. The conditions above, for geometrically equivalent Mach-Zehnder interferometers in the absence of difference of gravitational potential between each path of the array, lead us to the following constraint on the proper lengths when we rotate the array
\begin{eqnarray}
L_{1} &=& L_{2} \left( 1-gH/c^2 \right) + 2H, \nonumber \\
L'_1 &=& L'_2 \left( 1+gH/c^2 \right) + 2H.
\end{eqnarray}
As a consequence, the temporal delays become
\begin{eqnarray}
\Delta \tau_{\gamma_1,\gamma_2} &=& -gH L'_2/c^3,  \label{HgDeltaFinal1}\\
\Delta \tau_{\gamma'_1,\gamma'_2} &=& gH L'_2/c^3. \label{HgDeltaFinal2}
\end{eqnarray}
For photons that propagate in the same Mach-Zehnder interferometer the temporal delays become
\begin{eqnarray}
\Delta \tau_{\gamma_1,\gamma'_2} &=&  gH L'_2/c^3, \nonumber \\
\Delta \tau_{\gamma'_1,\gamma_2} &=&  -gH L'_2/c^3.
\end{eqnarray}
Consequently, it is now possible to detect two photons at different output ports of the same Mach-Zehnder interferometer.

Unfortunately, in presence of frequency dispersion and when the detection process is carried out with detectors that do not resolve frequency, the amount of entanglement is reduced. Thereby, Franson and Hugged interferometric arrays generate in presence of a weak gravitational field and frequency dispersion a partially entangled state. In this scenario, a non-maximal violation of the CHSH inequality is to be expected.
The effect of frequency dispersion can be best seen in the detection probabilities. In order to calculate the two-photon elapsed time detection probabilities, we assume that the detectors are insensitive to the frequency of the photons. Thereby, the action of the detectors is modelled by means of the projection operators
\begin{equation}
\widehat{P}_{i,\alpha} = \int d\omega  a^{\dagger}_{i,\alpha}(\omega)\vert{0}\rangle \langle{0}\vert a_{i,\alpha}(\omega),
\end{equation}
where $a_{i,\alpha}(\omega)$ denotes the annihilation operator of a single photon of frequency $\omega$ acting at the detector $i=\pm$ at the out port of the Mach-Zehnder interferometer to the left ($\alpha=1$) or to the right ($\alpha=2$). The probability for a detection at detector $(i,1)$ and a detection at detector $(j,2)$ is given by
\begin{equation}
p_{i,j} = \langle\Psi  \vert \widehat{P}_{i,1}\widehat{P}_{j,2} \vert{\Psi}\rangle, 
\end{equation}
which becomes
\begin{eqnarray}
p_{i,j} 
&=&  \frac{1}{4}\Big( 1 - (-1)^{\delta_{ij}} \int d\omega_1 d\omega_2 \vert f(\omega_1,\omega_2) \vert^2  \nonumber \\
&&  \cos(\omega_1  \Delta \tau_{\gamma_1\gamma_2} + \omega_2  \Delta\tau_{\gamma'_1\gamma'_2} + \alpha + \beta) \Big). 
\label{ProbFransonUniversal}
\end{eqnarray}
This turns out to be independent of the elapsed detection time $\Delta\tau$. It is, however, a function of the gravitational time delays $\Delta \tau_{\gamma_1\gamma_2}$ and $\Delta\tau_{\gamma'_1\gamma'_2}$, which are coupled trough the frequency of the photons to the frequency dispersion. In the case that all arms of the interferometric arrays are at the same gravitational potential, the gravitational time delays vanish and the elapsed detection time can be set to zero. Thereby, the probability $p_{i,j}$ becomes as Eq. (\ref{ProbFRHG-NORMAL}) with a maximal visibility, which is the known coincidence probability of the Franson and Hugged interferometric arrays.
 
Assuming that the distribution of frequencies for each photon is Gaussian, that is, $f(\omega_1,\omega_2)=f_1(\omega)f_2(\omega)$ with $f_i(\omega ) =   \exp\left(-\left( \omega - \omega_i \right)^2/ \left( 2 \sigma_{i}^2 \right)\right)/\left( \sigma_{i} \sqrt{2\pi} \right)$,
where $\sigma_i$ and $\omega_i$ are the width and the mean value of the Gaussian distribution, respectively, we can solve the integral entering in $p_{i,j}$ to obtain
\begin{eqnarray}
p_{i,j} &=& \frac{1}{4} \left( 1 - (-1)^{\delta_{ij}}V(\Delta \tau_{\gamma_1\gamma_2},\Delta \tau_{\gamma'_1\gamma'_2}) \right. \nonumber \\
&& \left. \cos \left[ \Delta \tau_{\gamma_1\gamma_2}\omega_1 + \Delta \tau_{\gamma'_1\gamma'_2} \omega_2 + \alpha + \beta \right] \right), \label{ProbGen}
\end{eqnarray}
where 
\begin{eqnarray}
V(\Delta \tau_{\gamma_1\gamma_2},\Delta \tau_{\gamma'_1\gamma'_2}) = \exp \left[ -\frac{1}{4}\left(\Delta \tau^{2}_{\gamma_1\gamma_2}\sigma_{1}^2 + \Delta \tau^{2}_{\gamma'_1\gamma'_2}\sigma_{2}^2 \right) \right]  \label{Visibility}
\end{eqnarray}
is the interferometric visibility of the two-photon detection process. These two expressions are valid for a Franson or Hugged array of any geometric configuration and in the presence of any weak gravitational field.

A further simplification of Eqs.\thinspace(\ref{ProbGen}) and (\ref{Visibility}) can be obtained by recalling that for a balanced Franson or Hugged array we have that
\begin{eqnarray}
\Delta\tau_{\gamma_1\gamma_2} &=& \Delta\tau_{\gamma'_1\gamma'_2}=\Delta\tau_{\gamma}\approx \frac{L'_2gH}{c^3}. \label{DeltaTauGen}
\end{eqnarray}
Thereby, we finally obtain for the two-photon elapsed time detection probability the expression
\begin{eqnarray}
p_{i,j} &=& \frac{1}{4} \left( 1 - (-1)^{\delta_{ij}}V(\Delta \tau_{\gamma})  \cos \left[ \Delta \tau_{\gamma}(\omega_1+\omega_2) + \alpha + \beta \right] \right),\nonumber \\
\label{finalProb}
\end{eqnarray}
with 
\begin{equation}
V(\Delta \tau_{\gamma}) = \exp\left[-\frac{1}{4}\Delta \tau^{2}_{\gamma}(\sigma_{1}^2 + \sigma_{2}^2)\right].
\end{equation}
We can identify two different regimes. If the condition $\Delta\tau^{2}_{\gamma}(\sigma^{2}_{1}+ \sigma^{2}_{2}) \ll 1$ holds, then the visibility is nearly maximal and the probability for elapsed detection oscillates harmonically according to the total phase shift given by $\Delta \tau_{\gamma}(\omega_1+\omega_2)$.
In the opposite regime, that is, $\Delta\tau^{2}_{\gamma}(\sigma^{2}_{1}+ \sigma^{2}_{2}) \gg 1$, the exponential decay of the visibility dominates and the harmonic oscillation tends to vanish. In this case, the exponential decay does not depend on the frequencies $\omega_1$ and $\omega_2$, but on the gravitational time delay and the width of the frequency distribution.

In the case that the gravitational temporal delay becomes negligible, i.e., $\Delta \tau_\gamma \rightarrow 0$, the probability of detection becomes $p_{ij} = \left(1 -(-1)^{\delta_{ij}}\cos(\alpha+\beta) \right)/4$, which is the known detection probability of the Franson and Hugged interferometric arrays.

\begin{figure}[t]
    \centering
    \includegraphics[width=0.9\textwidth]{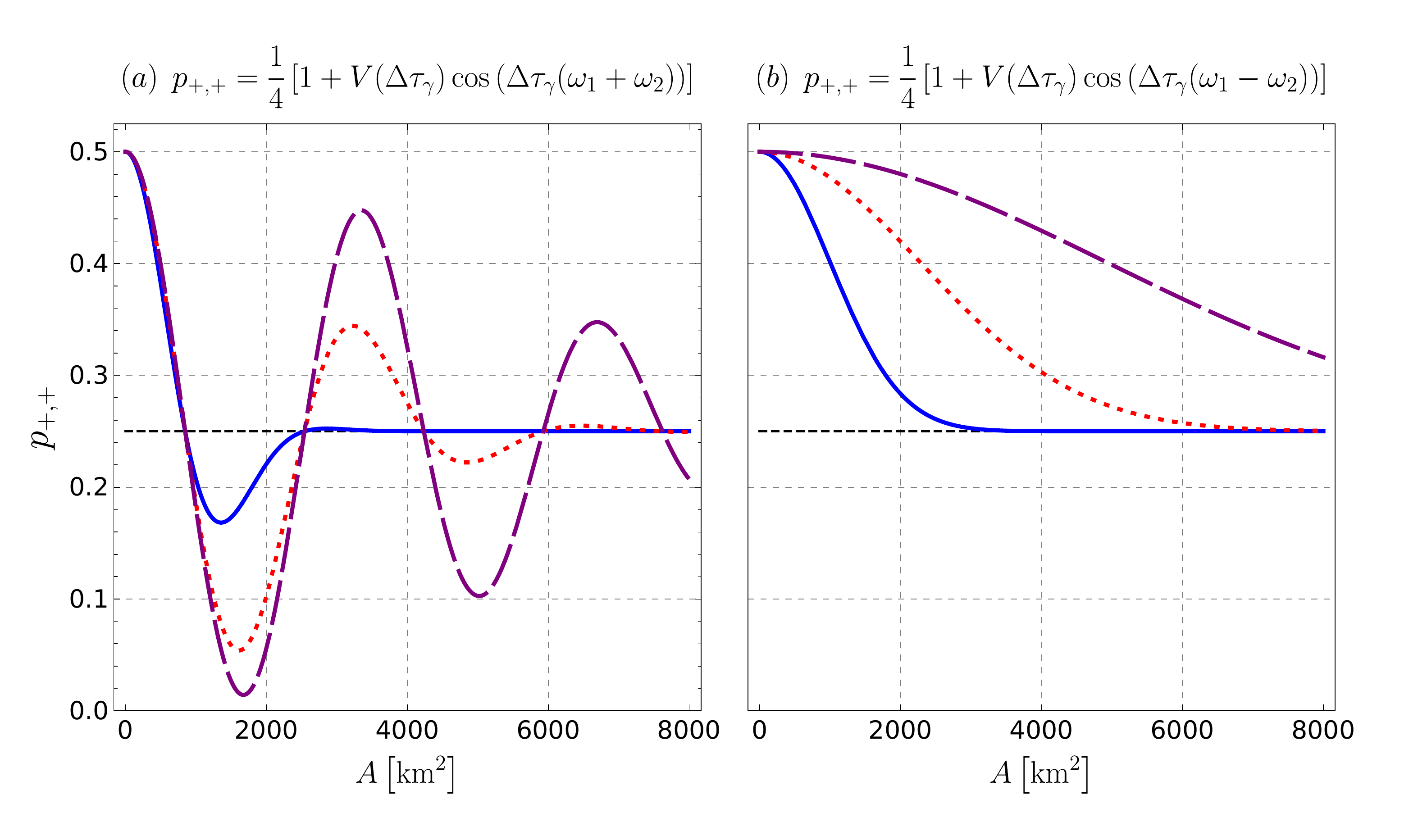}
    \caption{Elapsed time detection probability $p_{+,+}$, according to Eq.\thinspace(\ref{ProbGen}) as a function of the proper area $A=L'_2H$, and visibility $V(\Delta \tau_{\gamma}) =  \exp\left[-\Delta \tau^{2}_{\gamma}(\sigma_{1}^2 + \sigma_{2}^2)/4\right]$. (a) Franson and Hugged interferometric arrays with temporal delays $\Delta \tau_{\gamma} = L'_{2}gH/c^3$. (b) Geometrically identical, balanced and rotated Hugged array for the temporal delays given by Eqs.(\ref{HgDeltaFinal1}) and (\ref{HgDeltaFinal2}).  For both simulations $\alpha=\beta=0$, $\lambda_1$ = 806 nm, $\lambda_{2}$ = 706 nm, and $\delta \lambda_1 = \delta \lambda_2 = \delta \lambda$ with
$\delta \lambda= 161.2$ nm (purple dashed line), $\delta \lambda=322.4$ nm (red dotted line), and $\delta \lambda=644.8$ nm (blue continuous line). Black dashed line denotes a probability of detection of 1/4.}
    \label{fig:Bell1}
\end{figure}

In Fig.\thinspace\ref{fig:Bell1}(a) we display the behavior of the elapsed time detection probability $p_{+,+}$ according to Eq.\thinspace(\ref{finalProb}) as a function of the product $A=L_{\gamma'2}H$. Here we consider that twin photons are produced by an ultra-broadband laser source \cite{Dvoyrin2017, Nakajima2019, Gmachl2002}, where the value of the central wavelength is comparable to the value of the bandwidth. For each twin photon of the pair we consider wavelength distributions centred at   
 $\lambda_1 = 806$ nm and $\lambda_2 = 706$ nm with equal bandwidth $\delta \lambda_1 = \delta \lambda_2 =\delta \lambda$. We assume three different values of the width: $\delta \lambda = 161.2$ nm ($\sigma_1 = 4.724\times 10^{14}$ Hz and $\sigma_2 = 6.176\times 10^{14}$ Hz),  $\delta \lambda = 322.4$ nm ($\sigma_1 = 9.744\times 10^{14}$ Hz and $\sigma_2 = 1.286\times 10^{15}$ Hz), and $\delta \lambda = 644.2$ nm ($\sigma_1 = 2.224\times 10^{15}$ Hz and $\sigma_2 = 3.076\times 10^{15}$ Hz). The relation between the spectral width $\sigma$ and $\delta \lambda$ is given by $\sigma = 2\pi c(1/\lambda_{\rm min} - 1/\lambda_{\rm max})$, where $\lambda_{\rm max} = \lambda_0 + \delta \lambda/2$, $\lambda_{\rm min} = \lambda_0 - \delta \lambda/2$ , and $\lambda_0$ is the central wavelength of the wave packet. As is apparent from the figure, the larger the wavelength bandwidth $\delta \lambda$ the stronger is the damping of the harmonic oscillation.

Let us now assume a balanced and a geometrically equivalent Hugged interferometric array on an equipotential gravitational surface that is rotated 90º on the vertical. We also consider that $\omega_1=\omega_2$. In this case we have that the temporal delays entering in Eq.\thinspace(\ref{finalProb}) are given by $\Delta \tau_{\gamma_1\gamma_2}=-\Delta \tau_{\gamma'_1\gamma'_2} \approx -L_{2}H g/c^3 = - \Delta \tau_{\gamma}$. Consequently, the elapsed time two-photon detection probability becomes
\begin{equation}
p_{i,j} = \frac{1}{4} \left( 1 - (-1)^{\delta_{ij}}V(\Delta \tau_\gamma)\cos(\alpha + \beta) \right).
\label{HuggedBalanced}
\end{equation}
Thereby, the oscillatory behavior is no longer dictated by the gravitational temporal delay. The influence of a weak gravitational field manifests itself as a decrease in the visibility of $p_{i,j}$ only.

%\begin{figure}[t]%[H]
%    \centering
%    \includegraphics[width=0.5\textwidth]{FIG4V4.pdf} 
%    \caption{Elapsed time detection probability $p_{i,j}$, according Eq.\thinspace(\ref{finalProb}), for a geometrically equivalent and balanced Hugged interferometric array as a function of the proper area $A=L'_2H$ with $\alpha=\beta=0$, with wavelength $\lambda_1$ = 471.24 nm, $\lambda_{2}$ = 376.99 nm,
%$\sigma= 6283.19$ nm (purple dashed line), $\sigma=1346$ nm (red dotted line), and $\sigma=628.32$ nm (blue continuous line).}
%    \label{fig:Bell2}
%\end{figure}

 In Fig.\thinspace\ref{fig:Bell1}(b) we display the elapsed time detection probability of a balanced and geometrically equivalent Hugged interferometric array, according to Eq.\thinspace(\ref{finalProb}), considering nearly equal frequencies $\omega_1$ and $\omega_2$. In this case, the harmonic oscillation is almost suppressed and the probability is mainly dominated by the exponential decay, which is a function of the product between the gravitational time delay and the width of the Gaussian wave packets. In a single Mach-Zehnder interferometer in presence of a weak gravitational field and frequency dispersion, the detection probability also is given by the exponential decay of harmonic oscillation. In this case, however, it is not possible to isolate the exponential decay as in the case of a balanced Hugged interferometric array.

The elapsed time detection probability $p_{i,j}$ is a function of the gravitational time dilation. However, this result cannot be interpreted as a genuine test of General Relativity. A similar result can be obtained by considering light propagating in a Mach-Zehnder interferometer, where the spacetime is assumed to be flat and photons are coupled to the Newtonian gravitational potential provided that one considers an effective mass equal to the photon energy divided by $c^2$, as for a massive particle. The exponential decrease of visibility is also introduced in this case by means of frequency dispersion. The previous situation is similar to the case of the prediction of the gravitational redshift, which at the lowest order can also be obtained in the same way, since both effects are consequences of the gravitational time dilation. Furthermore, these effects are only sensitive to the first correction to the temporal component of the metric, and therefore do not depend, to the order of approximation here considered, e.g., on the $\gamma$ and $\beta$ post-Newtonian parameters (see for instance \cite{Will}).

\section{CHSH Inequality}
\label{SEC5}

In order to study the entanglement properties of the two-photon state generated by the Franson and Hugged interferometric arrays in presence of a weak gravitational field, we calculate the value of the functional $\Sigma$ of Eq.\thinspace(\ref{CHSH1}). We consider the case in which the light source at the interferometric arrays exhibits frequency dispersion. The use of $\Sigma$ to inspect the entanglement of the generated states is motivated by the lack of entanglement measures for non-gaussian multimode states.

The maximally entangled state $\vert \Psi\rangle$ of Eq.\thinspace(\ref{SSLL}) is obtained in the Franson array as a consequence of a post-selection procedure, where twin photons with elapsed detection times are neglected. Thereby, the state $\vert \Psi\rangle$ describes the fraction of photons that are detected in coincidence. A typical experiment aimed at measuring the functional $\Sigma$ would be based on the state $\vert \Psi\rangle$. However, it has been shown \cite{Aerts1999} that the neglected fraction of photons can be employed to support a hidden-variable model that simulates the quantum mechanical predictions for the experiment, that is, the Franson array is affected by a loophole. This does not exist if the local phases $\alpha$ and $\beta$ are varied in a time scale shorter than $\Delta L/c$. Nevertheless, we will employ the functional $\Sigma$ to study the influence of the weak gravitational field on the entanglement properties of the post-selected state generated by the Franson and Hugged arrays. The later is loophole-free and several methods have been proposed to eliminate the loophole affecting the Franson array, such as, for instance, polarization and energy-time hyperentanglement and time-bin entanglement. The latter is achieved by using a pulsed light source and replacing the beam splitters near the light source with small balanced Mach-Zehnder interferometers with fast and equal phase modulation.

\begin{figure}[t]
    \centering
    \includegraphics[width=0.55\textwidth]{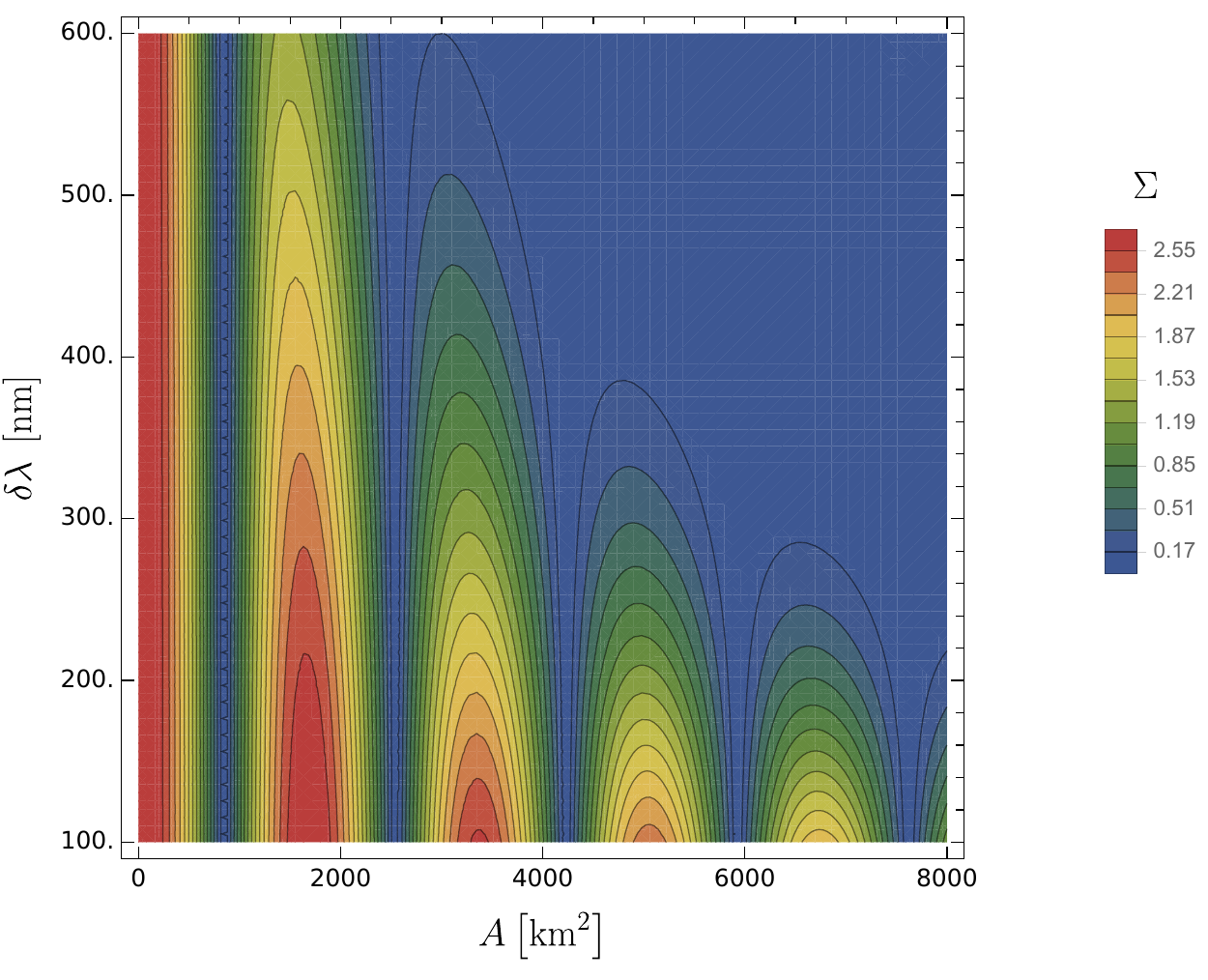}
    \caption{Value of the CHSH functional $\Sigma$ according Eq.\thinspace(\ref{CHSHfrhg}) for balanced Franson and Hugged interferometric arrays under condition Eq.\thinspace(\ref{Condition}), as a function of the proper area $A=L'_2H$ and the wave packet bandwidth $\delta \lambda$. We used $\left(\alpha,\beta,\alpha',\beta' \right) = \left(\pi/4,0, -\pi/4,-\pi/2\right)$, and $\lambda_1 = 806$ nm, and $\lambda_{2} = 706$ nm.}
    \label{fig:BellFRHG}
\end{figure}

\begin{figure}[t]
    \centering
    \includegraphics[width=0.9\textwidth]{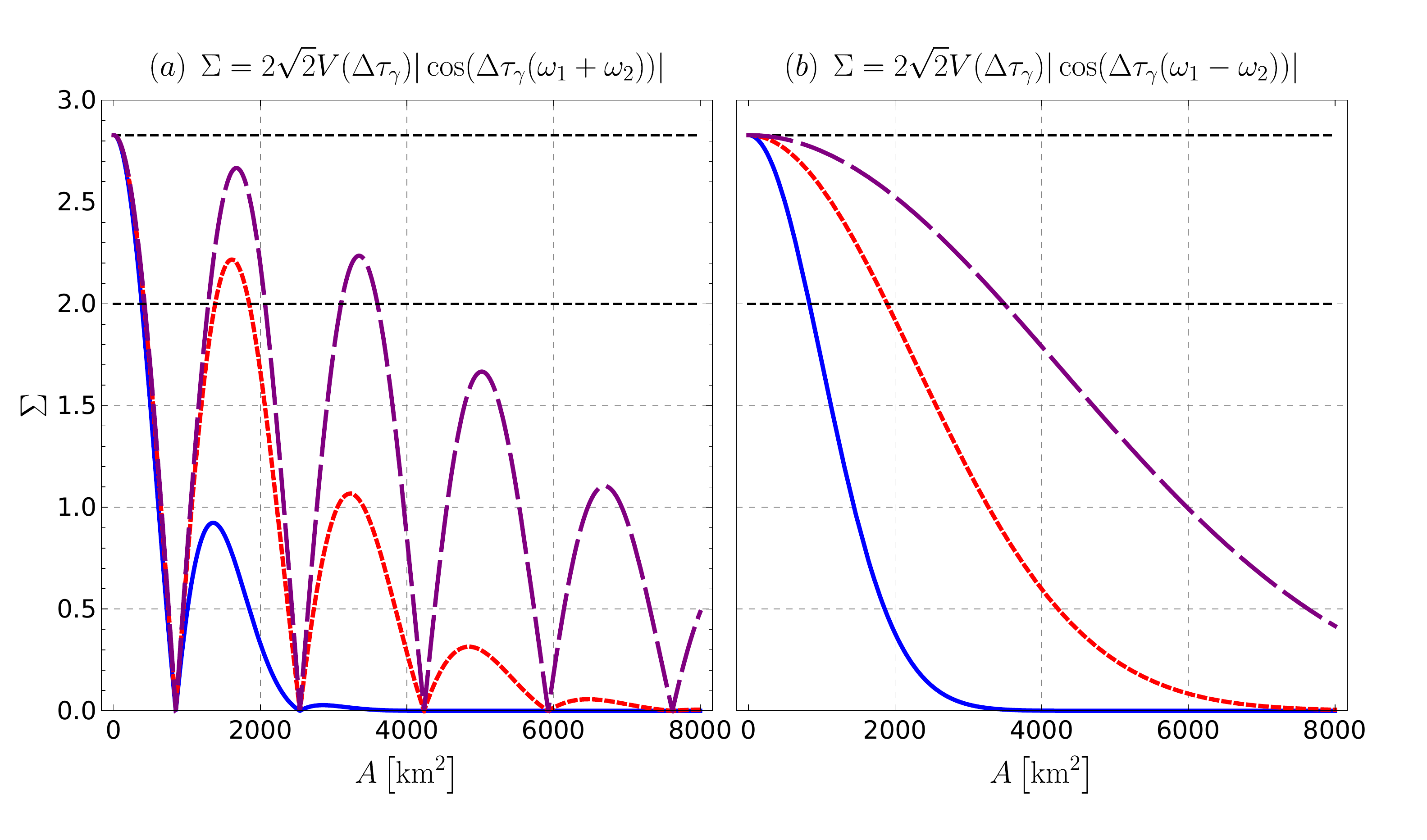}
    \caption{Value of $\Sigma$ for a balanced Hugged interferometric array as a function of the proper area $A=L'_2H$, elapsed proper time $\Delta \tau_{\gamma} = L'_{2}gH/c^3$, and visibility  $V(\Delta \tau_{\gamma}) = \exp\left[-\Delta \tau^{2}_{\gamma}(\sigma_{1}^2 + \sigma_{2}^2)/4\right]$. (a) Balanced Franson and Hugged arrays under condition Eq.\thinspace(\ref{Condition}) according to Eq.\thinspace(\ref{CHSHfrhg}). (b) Balanced, geometrically identical, and rotated Hugged array according to Eq.\thinspace(\ref{CHSHhUg}). For (a) and (b) we have: $\left(\alpha,\beta,\alpha',\beta' \right) = \left(\pi/4,0, -\pi/4,-\pi/2\right)$,  $\lambda_1$ = 806 nm, $\lambda_{2}$ = 706 nm, and $\delta \lambda_1 = \delta \lambda_2 = \delta \lambda$ with
$\delta \lambda= 161.2$ nm (purple dashed line), $\delta \lambda=322.4$ nm (red dotted line), and $\delta \lambda=644.8$ nm (blue continuous line). The horizontal black dashed line represents the maximal achievable value $2\sqrt{2}$ of the CHSH inequality.}
    \label{fig:BellHG}
\end{figure}

\begin{figure}[t]
    \centering
    \includegraphics[width=0.55\textwidth]{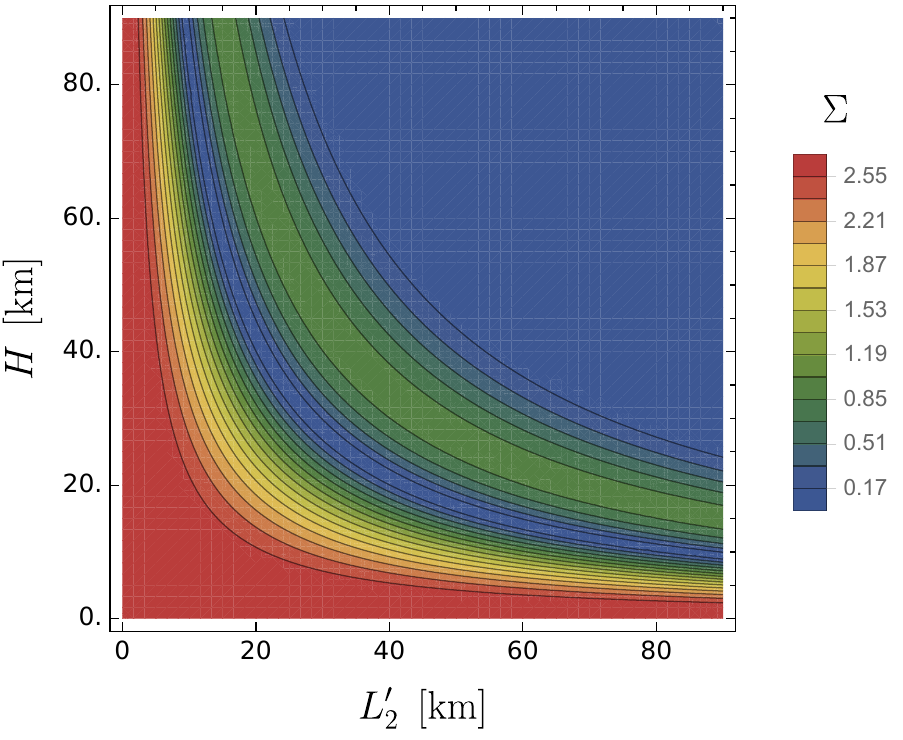}
    \caption{Value of $\Sigma$ for balanced Franson and Hugged interferometric arrays under condition Eq.\thinspace(\ref{Condition}) as a function of the proper length $L'_2$ and proper height $H$, according to Eq.\thinspace(\ref{CHSHfrhg}), considering $A= L'_{2} H$. We have chosen $\left(\alpha,\beta,\alpha',\beta' \right) = \left(\pi/4,0, -\pi/4,-\pi/2\right)$, $\lambda_{1} = 806$ nm, $\lambda_{2} = 706$ nm, and $\delta \lambda_1 = \delta \lambda_2 =644.8 $ nm.}
    \label{fig:BellFRHG-arms}
\end{figure}

Employing the elapsed time detection probability $p_{i,j}$, Eq.\thinspace (\ref{ProbFransonUniversal}), the expectation value $E(\alpha,\beta)$ of the dichotomic observable defined by phases $\alpha$ and $\beta$ becomes
\begin{eqnarray}
E\left(\alpha,\beta\right)& =& \int d\omega_1 d\omega_2 \left\vert f\left(\omega_1,\omega_2\right) \right\vert^2 \cos\left[ \omega_1 \Delta \tau_{\gamma_1,\gamma_2} + \omega_2 \Delta \tau_{\omega'_1,\omega'_2} + \alpha + \beta\right]. \nonumber \\
\end{eqnarray}
The resting expectation values are obtained by changing the value of the local phases. The CHSH functional $\Sigma$,  
assuming a Gaussian frequency distribution for each twin photon generated by the light source, and choosing the local phases as in the case of a maximally entangled Bell state, i.e., $\left( \alpha,\beta,\alpha',\beta' \right)=\left( \pi/4,0,-\pi/4,-\pi/2\right)$,  becomes 
\begin{eqnarray}
\Sigma &=& 2\sqrt{2}e^{-\frac{1}{4}\left( \Delta \tau_{\gamma_1,\gamma_2}^2 \sigma_{1}^2 + \Delta \tau_{\gamma'_1,\gamma'_2}^2 \sigma_{2}^2 \right)}|\cos \left( \omega_1 \Delta \tau_{\gamma_1,\gamma_2} + \omega_2 \Delta \tau_{\gamma'_1,\gamma'_2} \right)|. 
\end{eqnarray}
%This expression can be simplified under the assumption that $\alpha=\beta=0$, in which case we obtain
%\begin{equation}
%\Sigma=2\sqrt{2}|4p_{ii}-1|.
%\end{equation}
For a balanced Franson or a balanced Hugged interferometric array fulfilling the indistinguishability condition, the CHSH functional adopts the following  form
\begin{eqnarray}
\Sigma &=& 2\sqrt{2} e^{-\frac{1}{4}\Delta \tau^2_{\gamma} \left( \sigma_{1}^2 + \sigma_{2}^2 \right)}\left\vert  \cos\left( \Delta \tau_{\gamma} (\omega_1 + \omega_2)\right) \right\vert . \label{CHSHfrhg}
\end{eqnarray}
Figure (\ref{fig:BellFRHG}) illustrates the behaviour of the CHSH functional $\Sigma$ of Eq.\thinspace(\ref{CHSHfrhg}) in terms of the proper area $A=L'_2H$ and the width of the wave packet $\sigma$. When the arms of the Franson or Hugged interferometric arrays are at the same gravitational potential,  the functional $\Sigma$ leads to a maximum value of $2\sqrt{2}$, which is a violation of the CHSH inequality. However, in the presence of a weak gravitational field together with frequency dispersion, the CHSH functional $\Sigma$ exhibits a harmonic behavior with an amplitude that is exponentially damped (see Fig. \ref{fig:BellHG}(a)). As soon as the proper area $A$ is larger than $A_{*}=\sqrt{\ln 4} \left(c^3/g\sqrt{\sigma^{2}_{1} +\sigma^{2}_{2}} \right)$, it is not possible to violate the CHSH inequality. Let us note that the value of the upper bound $A_{*}$ is proportional to $(\sigma^{2}_1 + \sigma^{2}_2)^{-1/2}$ and, consequently, an increase of $\sigma_1$ and $\sigma_2$ implies that CHSH inequality will be violated in a narrower interval of proper area. Within this interval, the functional $\Sigma$ oscillates between values of $A$ which violate or not the CHSH inequality.
In Fig. \ref{fig:BellFRHG-arms} we show the functional $\Sigma$ as a function of $H$ and $L'_{2}$. As is apparent from this figure the CHSH inequality is violated for moderate values of the lengths $H$ and $L'_{2}$. For instance, for $H=L'_{2}=10$ [km] and a ultra-broad bandwidth  source like the one employed in \cite{Dvoyrin2017, Nakajima2019, Gmachl2002}, the value of $\Sigma$ is approximately 2.55, which is well above the classical upper bound of 2. An experimental realization of this gravitationally generated entanglement could consider surface stations for both the light source and the detectors, where photons are transmitted from the source to the detectors via satellites. On the light of recently performed experiments on the large-scale free-space propagation of photons \cite{Peng2005,Yin2017, Yin2012}, such an experiment might be feasible with current technology.

\begin{figure}[t]
    \centering
    \includegraphics[width=0.9\textwidth]{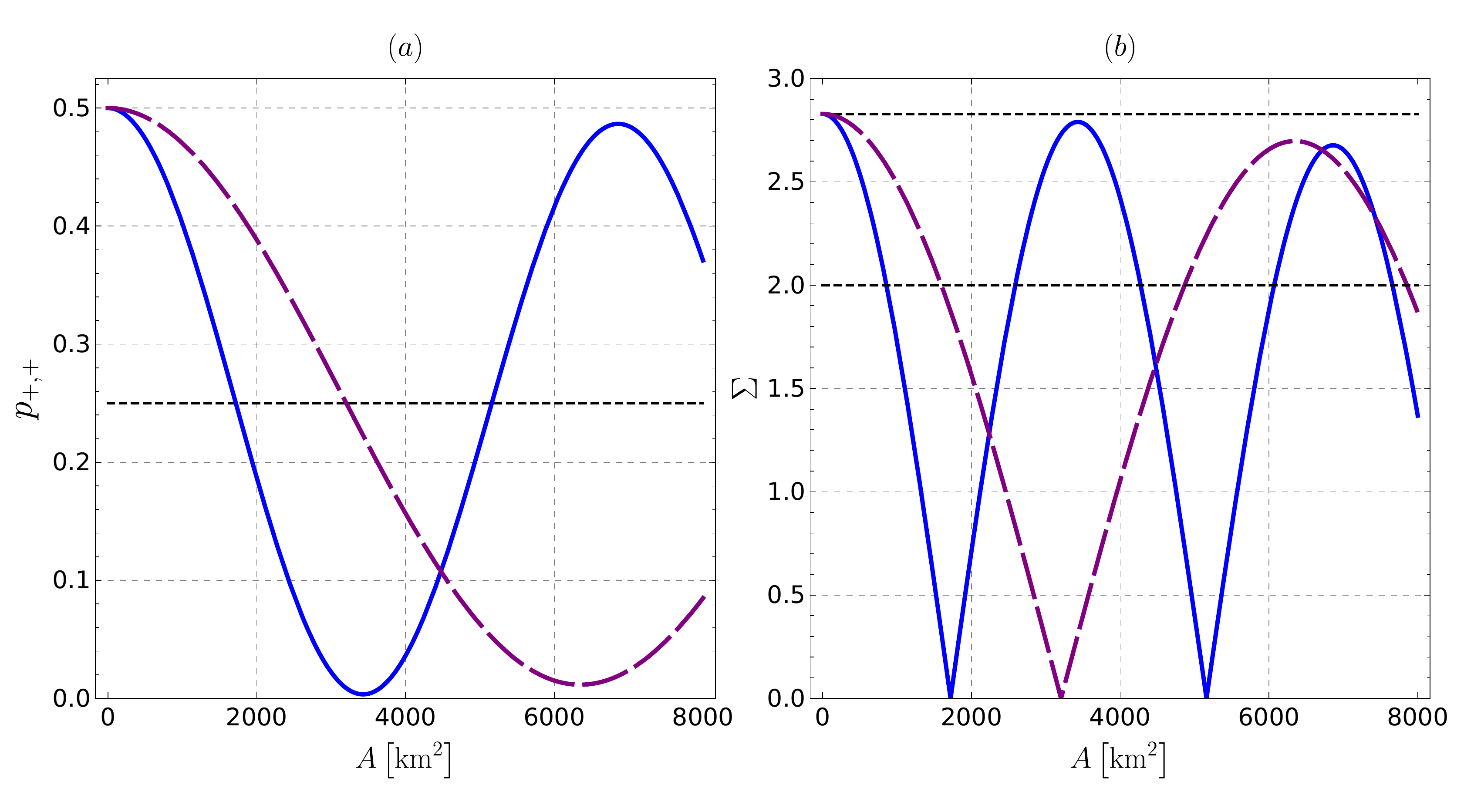}
    \caption{Elapsed time detection probability $p_{+,+}$ and $\Sigma$  as a function of the proper area $A=L'_2H$, with visibility $V(\Delta \tau_{\gamma}) =  \exp\left[-\Delta \tau^{2}_{\gamma}(\sigma_{1}^2 + \sigma_{2}^2)/4\right]$, for an ultra-broadband SPDC source \cite{Vanselow2019}. (a) \textit{ blue continuous line}: Elapsed time detection probability $p_{+,+}$ for a balanced Franson and Hugged interferometric arrays with temporal delays $\Delta \tau_{\gamma} = L'_{2}gH/c^3$. \textit{Purple dashed line}: $p_{+,+}$ for a geometrically identical, balanced and rotated Hugged array for the temporal delays given by Eqs.(\ref{HgDeltaFinal1}) and (\ref{HgDeltaFinal2}). (b) \textit{blue continuous line}: CHSH functional $\Sigma$ for a balanced Franson and Hugged interferometric arrays under condition (\ref{Condition}) and Eq. (\ref{CHSHfrhg}). \textit{Purple dashed line}: balanced, geometrically identical, and rotated Hugged array according to Eq.\thinspace(\ref{CHSHhUg}). For (a) and (b) the SPDC source employs a SLT crystal with the idler signal of $\lambda_1 = 3300$ nm, signal wave-packet with wavelength $\lambda_2 = 995$ nm, $\delta\lambda_1 = 370$  nm and $\delta\lambda_2 = 34$ nm. }
    \label{fig:Bell1REALSPDC}
\end{figure}

The oscillatory behavior displayed by the functional $\Sigma$ can be suppressed in the case of a balanced, geometrically equivalent, and rotated Hugged interferometric array, where we obtain
\begin{eqnarray}
\Sigma &=& 2\sqrt{2} e^{-\frac{1}{4}\Delta \tau^2_{\gamma} \left( \sigma_{1}^2 + \sigma_{2}^2\right)} \left\vert\cos\left( \Delta \tau_{\gamma} (\omega_1 - \omega_2)\right)\right\vert .\label{CHSHhUg}
\end{eqnarray}
Clearly, the choice $\omega_1=\omega_2$ eliminates the harmonic behavior.  This is illustrated in Fig.\thinspace\ref{fig:BellHG}(b) for slightly different values of $\omega_1$ and $\omega_2$. A similar result can be obtained without constraining the values of the frequencies. In order to do this the local phases can be measured with respect to the temporal delays for each photon along each path in the interferometric arrays, that is, $\alpha= \varphi_a - \omega_1 \Delta \tau_{\gamma_1,\gamma_2}$ and $\beta = \varphi_b  - \omega_1 \Delta \tau_{\gamma'_1,\gamma'_2}$ (with $a, b=1, 2$), where $\varphi_a$ and $\varphi_b$ are chosen such as the difference between these local phases and the gravitational time delays correspond to the optimal values of the local phases to maximally violate the CHSH inequality. In this case the CHSH inequality becomes
\begin{eqnarray}
\Sigma &=& e^{-\frac{1}{4}\left( \Delta \tau_{\gamma_1,\gamma_2}^2 \sigma_{1}^2 + \Delta \tau_{\gamma'_1,\gamma'_2}^2 \sigma_{2}^2 \right)} 
 \left[ \cos\left( \varphi_1 + \varphi_2 \right) \right. \nonumber \\
 && \left. + \cos\left( \varphi_1 + \varphi'_2 \right) + \cos\left( \varphi'_1 + \varphi_2 \right) - \cos\left( \varphi'_1 + \varphi'_2 \right) \right]. \nonumber \\ \label{CHSHopVal1}
\end{eqnarray}
Considering the optimal local phases $(\varphi_1,\varphi_2,\varphi'_1,\varphi'_2) = (\pi/4,0,-\pi/4,-\pi/2)$ the CHSH function (\ref{CHSHopVal1}) becomes
\begin{eqnarray}
\Sigma &=&  2\sqrt{2} e^{-\frac{1}{4}\left( \Delta \tau_{\gamma_1,\gamma_2}^2 \sigma_{1}^2 + \Delta \tau_{\gamma'_1,\gamma'_2}^2 \sigma_{2}^2 \right)},
\end{eqnarray}
which holds for Franson and Hugged interferometric arrays. The previous expression exhibits the exponential decay without the harmonic modulation. This, without requiring a balanced Hugged interferometric array or constraining the frequencies of the twin photons.

%\begin{figure}[t]
%    \centering
%    \includegraphics[width=0.5\textwidth]{FIG7.pdf}
%    \caption{Value of $\Sigma$ for a balanced Hugged interferometric array as a function of the proper area $A=L'_2H$, according Eq.\thinspace(\ref{CHSHhUg}), for $\left(\phi_{A},\phi_{B},\phi_{A'},\phi_{B'} \right) = \left(\pi/4,0, -\pi/4,-\pi/2\right)$, with wave-length $\lambda_{1} = 471.239$ nm, $\lambda_{2} = 376.991$ nm,
%$\sigma= 6283.19$ nm (purple dashed line), $\sigma=1346.4$ nm (red dotted line), and $\sigma=628.319$ nm (blue continuous line). The horizontal black dashed line represents the maximal achievable value of the CHSH inequality.}
%    \label{fig:BellHG}
%\end{figure}

Franson and Hugged interferometric arrays are configured on a gravity equipotential surface to achieve the maximal value $2\sqrt{2}$ of the functional $\Sigma$ allowed by the laws of Quantum Mechanic \cite{Lima2010, Cuevas2013}, which entails a violation of the CHSH inequality. The presence of a weak gravitational field affecting the propagation paths in Franson and Hugged interferometric arrays introduces a gravitational time delay. Thereby, simultaneously generated twin photons arrive to detectors at different times. This, however, does not preclude the generation of maximally entangled states or the violation of the CHSH inequality. This scenario changes abruptly if we allow for a light source exhibiting frequency dispersion. Assuming detectors that do not resolve frequency, the detection probabilities must be averaged over the two-photon frequency distribution of the source. This leads to a loss of the two-photon interferometric visibility, which is characterized by the exponential decay of the detection probabilities with the characteristic proper area of the Mach-Zehnder interferometers. This, in turn, leads to a functional $\Sigma$ characterized by the same exponential decay. Consequently, for Hugged and Franson interferometric arrays in the presence of a weak gravitational field, the value achieved by the functional $\Sigma$ might become smaller than the classically allowed upper bound of 2.

In the Newtonian limit, there is no difference of proper time in the arrival of the photon to the detectors. Then, the CHSH inequality only exhibits a harmonic behavior, which resembles the optical COW experiment \cite{Rideout2012} that involves a single Mach-Zehnder interferometer.
 
So far, we have considered a quantum description of the light propagated inside the interferometric arrays. Instead, if we consider a classical description of the light, then the functional $\Sigma$ (for further details, see \cite{Marcin2018, Ou1990}) adopts the form
\begin{eqnarray}
\Sigma_{\rm class} &=& 2\frac{\sqrt{2}}{4} e^{-\frac{1}{4}\left( \Delta \tau_{\gamma_1,\gamma_2}^2 \sigma_{1}^2 + \Delta \tau_{\gamma'_1,\gamma'_2}^2 \sigma_{2}^2 \right)} \Bigg| \cos \left( \omega_1 \Delta \tau_{\gamma_1,\gamma_2} + \omega_2 \Delta \tau_{\gamma'_1,\gamma'_2} \right) \Bigg|,
\end{eqnarray}
which can never exceed the classical bound of 2.

In the previous paragraphs we have studied the behavior of the elapsed detection probability and the CHSH functional by means of numerical simulations. These consider values of the wavelength and the wavelength width which are comparable, which is typical of ultra-broadband light sources. The values have been chosen such that the exponential dampening of the visibility can be  observed within realistic values of the area of the interferometric arrays. In Fig.\thinspace\ref{fig:Bell1REALSPDC} we consider values of wavelengths $\lambda_1$ and $\lambda_2$ and widths $\delta\lambda_1$ and $\delta\lambda_2$, respectively, which can be generated with an ultra-broadband SPDC source currently available \cite{Vanselow2019}. As can be appreciated in Fig.\thinspace\ref{fig:Bell1REALSPDC}, the exponential decay of the visibility is less pronounced within the studied area values. It is also clear that there are area values that allow the violation of the CHSH inequality and, consequently, the generation of entanglement through the gravitational time delay, even in the presence of frequency dispersion and for balanced Franson and Hugged interferometric arrays.

\section{Conclusions}
\label{SEC6}

We have studied the influence of a weak gravitational field on the generation of two-photon energy-time entangled states by means of large scale Franson and Hugged interferometric arrays. In absence of gravity, two simultaneously generated photons can be detected in coincidence, which indicates propagation along optical paths of equal length, or separated by a time interval, which indicates propagation along optical paths of different lengths. Simultaneously detected photons are described by a maximally entangled state given by the superposition of separable, mutually orthogonal states that describe the propagation of two photons along optical paths of equal lengths. Time dilation caused by the gravitational field introduces phase shifts in the two-photon detection probabilities such that a maximally entangled state becomes associated with elapsed time detections. The phase shift is given by the gravitational time delay times the photon frequency. Consequently, the elapsed time detection probabilities become cosine functions of the frequency of the photons generated by the light source. This is not a problem as long as the light source is monochromatic. Otherwise, if the light source exhibits frequency dispersion, the amplitude of the harmonic oscillation, characteristic of the elapsed time detection probabilities, decreases exponentially. In other words, the two-photon interference visibility of the arrays is reduced. This loss of visibility affects the generated two-photon state in such a way that the CHSH inequality cannot be violated if the proper area $A$ of the Mach-Zehnder interferometers is larger than or equal to $A_{*}=\sqrt{\ln 4} \left(c^3/g\sqrt{\sigma^{2}_{1} +\sigma^{2}_{2}} \right)$, in which case the system can be described by a classical theory. Thus, Franson and Hugged interferometric arrays with $A<A_*$ generate entangled states.

An interesting finding is that gravity can help to generate an entangled state. We considered a Franson interferometric array placed on a surface of equal gravitational potential and assume that its Mach-Zehnder interferometers are balanced and geometrically identical. This can be achieved by adding delay lines along the propagation paths of the photons, for example, with the help of fiber optic loops. Then all optical paths have the same length. Thereby, it is not possible to distinguish among pairs of paths since they are all associated with coincidence detections. In this case, it is not possible to create an entangled state. If we now rotate the interferometric array, in such a way that the arms of the interferometers are at different gravitational potentials and the Mach-Zehnder interferometers stay balanced and geometrically identical, in terms of its proper lengths, then photons following paths at different gravitational potentials experience a gravitational time delay. This leads to an elapsed time between detections. Photons following paths at equal gravitational potential are still detected in coincidence. Thereby, it is possible to distinguish among trajectories, perform the post-selection, and generate a maximally entangled state. In this scenario, we state that a weak gravitational field allows to generate a maximally entangled state. It is interesting to note that the post-selection loophole closed in the Hugged array, can also be closed in the Franson array considering the use of fast phase shifters \cite{Vedovato2018}.

The role of energy-time entanglement and the coincidence of events for the Franson and Hugged interferometer could be extended to clocks at different gravitational potentials in order to understand how two participants with different clocks affect the CHSH inequality. Other feasible extensions are the inclusion of the weak gravitational field generated by a rotating mass (for instance, following the results by Kolhurs {\it et al.}  \cite{Kohlrus2017}) and also to consider satellites with relative velocities, which require to take into account the effects of the Doppler shift and/or the possible ways to eliminate them (see the recent proposal in ref. \cite{Terno2019}).
\ack
This work was supported by the Millennium Institute for Research in Optics (MIRO) and FONDECYT Grant 1180558. M.R.-T. acknowledges support by CONICYT scholarship 21150323.

\appendix
\section{Phase shift in a Hugged interferometric array}\label{A1}
Here, we describe the phase shift produced in a Hugged interferometer (see fig. \ref{Hugged}).
The proper lengths of paths $\gamma_1$ and $\gamma'_1$ and the horizontal segments of paths $\gamma_2$ and $\gamma'_2$ are given by the following expressions
\begin{eqnarray}
L_{1} &\approx & \left( 1 -\frac{\phi(R+h)}{c^2} \right)\Delta x_{_1}, \\
L'_1 &\approx & \left( 1 -\frac{\phi(R+h)}{c^2} \right)\Delta x_{'_1},\\
L_{2} &\approx & \left( 1 -\frac{\phi(R)}{c^2} \right)\Delta x_{_2},\\
L'_2 &\approx & \left( 1 -\frac{\phi(R+2h)}{c^2} \right)\Delta x_{'_2},
\end{eqnarray}
while the proper length of the vertical segments is given by
\begin{equation}
H \approx \left( 1 -\frac{\phi(R)}{c^2} \right)h.
\end{equation}
The intervals of coordinate time can be obtained in a similar way. For the path $\gamma_1$ 
\begin{eqnarray}
\Delta t_{\gamma_1} &=& \left(1-2\frac{\phi(R+h)}{c^2} \right) \frac{\Delta x_{_1}}{c} 
\nonumber\\
&\approx & \left(1-\frac{\phi(R+h)}{c^2} \right) \frac{L{\gamma_1}}{c}.
\nonumber \\ 
\end{eqnarray}
For the path $\gamma'_1$
\begin{eqnarray}
\Delta t_{\gamma'_1} \approx \left(1-\frac{\phi(R+h)}{c^2} \right) \frac{L{\gamma'_1}}{c}.
\end{eqnarray}
For the horizontal part of the path $\gamma_2$ we have
\begin{eqnarray}
\Delta t^{H}_{\gamma_2} &=& \left(1-2\frac{\phi(R)}{c^2} \right) \frac{\Delta x_{_2}}{c} 
\nonumber\\
&\approx&  \left(1-\frac{\phi(R)}{c^2} \right) \frac{L_{2}}{c}.
\end{eqnarray}
The  coordinate time of the vertical part of the path $\gamma_2$ is given by
\begin{eqnarray}
\Delta t_{\gamma_2}^{V} &=& \frac{1}{c}\int_{R}^{R+h} \left( 1 -2 \frac{\phi(r)}{c^2} \right)dr \nonumber \\
%& \approx & \frac{1}{c}\int_{R}^{R+h} \left( 1 -2 \frac{\phi(R)}{c^2} -2 gr \right)dr \nonumber \\
& \approx & \frac{h}{c}\left( 1 -2 \frac{\phi(R)}{c^2} \right) \approx \frac{H}{c}\left(1-\frac{\phi(R)}{c^2} \right).
\end{eqnarray}
This is similar to the temporal coordinate for the vertical part of the path $\gamma'_2$. The coordinate time for the horizontal part of the path $\gamma'_2$ is
\begin{eqnarray}
\Delta t^{H}_{\gamma'_2} &=& \left(1-2\frac{\phi(R+2h)}{c^2} \right) \frac{\Delta x_{'_2}}{c} 
\nonumber\\
&\approx& \left(1 -\frac{\phi(R+2h)}{c^2}\right)\frac{L'_2}{c}.
\end{eqnarray}
The interval of proper time for the path $\gamma_1$ becomes
\begin{eqnarray}
\Delta \tau_{\gamma_1} &=& \left(1 + \frac{\phi(R + h)}{c^2} \right) \Delta t_{\gamma_1} \nonumber \\
& \approx & \left(1 + \frac{\phi(R+ h)}{c^2} \right)\left(1 - \frac{\phi(R+ h)}{c^2} \right)\frac{L_{1}}{c} \nonumber \\
& \approx & \frac{L_{1}}{c}.
\end{eqnarray}
Analogously, for the path $\gamma'_1$ we have
\begin{eqnarray}
\Delta \tau_{\gamma'_1} &\approx & \frac{L'_1}{c}.
\end{eqnarray}
For the path $\gamma_2$ we obtain
\begin{eqnarray}
\Delta \tau_{\gamma_2} &\approx & \left(1 + \frac{\phi(R + h)}{c^2} \right) \left(\Delta t^{H}_{\gamma_2} + 2 \Delta t^{V}_{\gamma_2} \right) \nonumber \\
& \approx & \left( 1 + \frac{\phi(R+h)}{c^2} \right) \left( \left(1-\frac{\phi(R)}{c^2} \right)\frac{L_{2}}{c^2} \right. \nonumber \\
&& \left. + 2 \left( 1 -\frac{\phi(R)}{c^2} \right)\frac{H}{c} \right) \nonumber \\
& \approx & \left(1 + \frac{g h}{c^2} \right)\frac{L'_2}{c} + \frac{2H}{c}\left(1 + \frac{gh}{c^2} \right) \nonumber \\
& \approx & \left(1 + \frac{g H}{c^2} \right)\frac{L_{2}}{c} + \frac{2H}{c},
\end{eqnarray}
where we have used $h \approx H (1+ \phi(R)/c^2)$ and an approximation to order $O(H^2)$.
Analogously, for the interval of proper elapsed time along the path $\gamma'_2$
\begin{eqnarray}
\Delta \tau_{\gamma'_2}  &\approx & \left(1 + \frac{\phi(R + h)}{c^2} \right) \left(\Delta t^{H}_{\gamma'_2} + 2 \Delta t^{V}_{\gamma'_2} \right) \nonumber \\
& \approx & \left( 1 + \frac{\phi(R+h)}{c^2} \right) \left( \left(1-\frac{\phi(R+ 2 h)}{c^2} \right)\frac{L'_2}{c^2} \right. \nonumber \\
&& \left. + 2 \left( 1 -\frac{\phi(R)}{c^2} \right)\frac{H}{c} \right) \nonumber \\ 
& \approx & \left(1 - \frac{g H}{c^2} \right)\frac{L'_2}{c} + \frac{2H}{c}.
\end{eqnarray}
Therefore, the difference of arrival time is 
\begin{eqnarray}
\Delta \tau_{\gamma_1,\gamma_2} &=& \Delta \tau_{\gamma_1} - \Delta \tau_{\gamma_2} \nonumber \\
& \approx & \frac{L_{1}}{c} - \frac{L_{2}}{c}\left(1 + \frac{gH}{c^2} \right) -2\frac{H}{c},  \label{taugamma12}
\end{eqnarray}
where we have used the following  constraints $L_{1} \approx L'_1 + c\Delta \tau$, $L_{2} \approx L'_2\left( 1 - 2gH/c^2 \right) + c \Delta \tau \left( 1 - gH/c^2 \right)$ and  $L'_1 = L'_2 + 2H$.

As in the case of the Franson array, to generate a two-photon entangled state, we impose the condition Eq.\thinspace(\ref{Condition}), which guarantees the indistinguishability of two twin photons traveling along paths $(\gamma_1,\gamma'_1)$ from those travelling along paths $(\gamma_2,\gamma'_2)$. In terms of the proper lengths associated to these paths, we obtain the equations
\begin{eqnarray}
L_{1}&\approx&L'_1+c\Delta\tau,
\\
L_{2}&\approx&L'_2(1-\frac{2gH}{c^2})+c\Delta\tau(1-\frac{gH}{c^2}).
\end{eqnarray}
These allow us to obtain the temporal delay of twin photons propagating along paths $(\gamma_1,\gamma_2)$ and $(\gamma'_1,\gamma'_2)$, that is,
\begin{eqnarray}
\Delta\tau_{\gamma_1\gamma_2}&\approx&\frac{L'_1}{c}-\frac{L'_2}{c}(1-\frac{gH}{c^2})-\frac{2H}{c},
\\
\Delta\tau_{\gamma'_1\gamma'_2}&=& \Delta\tau_{\gamma_1\gamma_2}.
\end{eqnarray}
We can now impose the condition Eq.\thinspace(\ref{NOGEO}), which removes the geometric terms entering in $\Delta\tau_{\gamma_1\gamma_2}$. Thereby, we obtain
\begin{equation}
\Delta \tau_{\gamma_1\gamma_2}=\Delta \tau_{\gamma'_1\gamma'_2}\approx \frac{L'_2 g H}{c^3}.
\end{equation}
The phase shift becomes
\begin{eqnarray}
\Delta \varphi &=& \omega \Delta \tau_{\gamma_1\gamma_2} = \omega \Delta \tau_{\gamma'_1\gamma'_2}  \approx  \frac{\omega L'_2 g H}{c^3}.
\end{eqnarray}

Let us now assume that the Hugged interferometric array is such that $L'_1=L'_2+2H$ and $L_{1}=L_{2}+2H$ hold. In this case, the difference of proper times measured by a clock at the detectors for the photon that moves through paths $\gamma_1$ and $\gamma_2$ becomes
\begin{equation}
\Delta \tau_{\gamma_1\gamma_2} \approx -\frac{L_{2}H g}{c^3}.
\label{ProperTimeHGgamma1gamma2}
\end{equation} 
Analogously, for a photon that moves through paths $\gamma'_1$ and $\gamma'_2$ we obtain 
\begin{equation}
\Delta \tau_{\gamma'_1\gamma'_2} \approx \frac{L'_2H g}{c^3}.\label{ProperTimeHGgamma1pgamma2p}
\end{equation} 
There are two differences between the elapsed proper time of the photon that propagates along paths $(\gamma_1, \gamma_2)$ with respect to the photon that moves along the paths $(\gamma'_1, \gamma'_2)$: the sign and the proper lengths $L_{2}$ and $L'_2$. The difference in the sign comes from the fact that photons arrive first at the detector on the right through the path $\gamma_2$ respect to a clock at $R+h$ and it will arrive later to the detector at the left through the path $\gamma_1$. The difference between the lengths (or the proper area $A=L\times H$) of the horizontal segments of paths $\gamma_2$ and $\gamma'_2$ can be calibrated to be $L_{2} = L'_2$ such that the Mach-Zehnder interferometers have the same effective proper area. Since the proper lengths $L_{2}$ and $L'_2$ are equal, the previous conditions demand that $L_{1} = L'_1$, which indicates that the Mach-Zehnder interferometers are balanced and have the same characteristic proper lengths. In this case we have that $\Delta\tau_{\gamma_1}=\Delta\tau_{\gamma'_1}$ and the temporal delay $\Delta\tau_{\gamma_1\gamma'_1}$ vanishes. However, the temporal delay involving paths $\gamma_2$ and $\gamma'_2$ becomes
\begin{equation}
\Delta\tau_{\gamma_2\gamma'_2} \approx 2\frac{L'_2gH}{c^3},
\end{equation}
which is twice the gravitational temporal delay entering in $\Delta \tau_{\gamma_1\gamma_2}$ and $\Delta \tau_{\gamma'_1\gamma'_2}$. In this case, twin photons propagating along paths $(\gamma_1,\gamma'_1)$ can be distinguished from twin photons propagating along paths $(\gamma_2,\gamma'_2)$ by means of the gravitational temporal delay between detections. 

Let us note that we can obtain a similar result by performing an upside-down rotation in one arm of the Franson array. However, the advantage of the Hugged array comes from the fact that it lacks the post-selection loophole characteristic of the Franson array \cite{Aerts1999, Cabello2009}.

\section{Probability of detection for Franson and Hugged interferometric arrays}\label{A2}
In this appendix, we obtain the various detection probabilities and the CHSH inequality for the Franson and Hugged interferometric arrays. We consider the bipartite state of Eq.\thinspace(\ref{InitialStateFRHG}). 

On both interferometric arrays, each photon undergoes the action of a Mach-Zehnder interferometer, which is given by the product of operators $U_{bs,i}U_{g,i}U_{bs,i}$. The operator $U_{bs,i}$ represents the action of a balanced beam splitter at input and output ports of a Mach-Zehnder interferometer. The operator $U_{g,i}$ represents the path-dependent temporal evolution of a photon inside the interferometer. The annihilation operators $a(\omega_1)$ and $a(\omega_2)$ are transformed by the operators $U_{g,i}U_{bs,i}$ into the linear combinations
\begin{eqnarray}
\frac{1}{\sqrt{2}}\left( a_{\gamma_1}(\omega_1) e^{i \omega_1 \Delta \tau_{\gamma_1}} + a_{\gamma_2}(\omega_1) e^{i \omega_1 \Delta \tau_{\gamma_2}} e^{i\alpha} \right), 
\\
\frac{1}{\sqrt{2}}\left( a_{\gamma'_1}(\omega_2) e^{i \omega_2 \Delta \tau_{\gamma'_1}} + a_{\gamma'_2}(\omega_2) e^{i \omega_2 \Delta \tau_{\gamma'_2}} e^{i\beta} \right), 
\label{Aes}
\end{eqnarray}
respectively. The operator $a_{\gamma_i}$ ($a_{\gamma'_i}$) describes the annihilation of a photon in the mode defined by path $\gamma_i$ ($\gamma'_i$). In Eq.\thinspace(\ref{Aes}), the evolution operator introduces a temporal dependence. This is given by the path-dependent proper times $\Delta \tau_{\gamma_i}$ and $\Delta \tau_{\gamma'_i}$ that are measured by a clock at the site of the detectors. We have also considered the controllable local phases $\alpha$ and $\beta$. 
 
Annihilation operators $a_{\gamma_i}$ and $a_{\gamma'_i}$ are related with the annihilation operators $a_{i,\alpha}$, which describe the modes after the output ports of the Mach-Zehnder interferometer, by the expressions
\begin{eqnarray}
a_{\gamma_1} = \frac{1}{\sqrt{2}} \left( a_{1,1}(\omega_1) + a_{1,2}(\omega_1) \right),  \\
a_{\gamma_2} = \frac{1}{\sqrt{2}} \left( a_{1,1}(\omega_1) - a_{1,2}(\omega_1) \right),  \\
a_{\gamma'_1} = \frac{1}{\sqrt{2}} \left( a_{2,1}(\omega_2) + a_{2,2}(\omega_2) \right),  \\
a_{\gamma'_2} = \frac{1}{\sqrt{2}} \left( a_{2,1}(\omega_2) - a_{2,2}(\omega_2) \right). 
\end{eqnarray}
Thereby, the initial two-photon wave packet $|\psi\rangle$ is transformed by the interferometric arrays into the state $|\psi'\rangle=U_{bs,1}U_{g,1}U_{bs,1}U_{bs,2}U_{g,2}U_{bs,2}|\psi\rangle$, which is given by
\begin{eqnarray}
\vert \psi' \rangle &=& \frac{1}{4} \int \int d\omega_1 d\omega_2 f(\omega_1,\omega_2) \left[ \left( a_{1,1}^{\dagger}(\omega_1) + a_{1,2}^{\dagger}(\omega_1) \right)	  \right. \nonumber \\
&& \left. \left( a_{2,1}^{\dagger}(\omega_2) + a_{2,2}^{\dagger}(\omega_2) \right) e^{-i \omega_1 \Delta \tau_{\gamma_1} - i \omega_2 \Delta \tau_{\gamma'_1}} \right. \nonumber \\
&& \left. +  \left( a_{1,1}^{\dagger}(\omega_1) - a_{1,2}^{\dagger}(\omega_1) \right)\left( a_{2,1}^{\dagger}(\omega_2) + a_{2,2}^{\dagger}(\omega_2) \right) \right. \nonumber \\
&& \times \left. e^{-i \omega_1 \Delta \tau_{\gamma_2} - i \omega_2 \Delta \tau_{\gamma'_1}} e^{i \alpha} \right. \nonumber \\
&& \left. + \left( a_{1,1}^{\dagger}(\omega_1) + a_{1,2}^{\dagger}(\omega_1) \right) \left( a_{2,1}^{\dagger}(\omega_2) - a_{2,2}^{\dagger}(\omega_2) \right) \right. \nonumber \\
&& \left. \times  e^{-i \omega_1 \Delta \tau_{\gamma_1} - i \omega_2 \Delta \tau_{\gamma'_2}} e^{i \beta} \right. \nonumber \\
&& \left. + \left( a_{1,1}^{\dagger}(\omega_1) - a_{1,2}^{\dagger}(\omega_1) \right) \left( a_{2,1}^{\dagger}(\omega_2) - a_{2,2}^{\dagger}(\omega_2) \right) \right. \nonumber \\
&& \left. \times  e^{-i \omega_1 \Delta \tau_{\gamma_2} - i \omega_2 \Delta \tau_{\gamma'_2}} e^{i ( \alpha + \beta )}\right]\vert{0}\rangle. 
\label{StateBeforeSelection} 
\end{eqnarray}
Now, we impose the indistinguishability of paths $(\gamma_1,\gamma'_1)$ and $(\gamma_2,\gamma'_2)$, that is, $\Delta\tau_{\gamma_1}-\Delta\tau_{\gamma'_1}=\Delta\tau$ and $\Delta\tau_{\gamma_2}-\Delta\tau_{\gamma'_2}=\Delta\tau$. 
In order to carry out the post-selection process, we apply the conditions to enforce indistinguishability. With these conditions the proper time becomes $\Delta \tau_{\gamma_1} = \Delta \tau_{\gamma'_1} + \Delta \tau$ and $\Delta \tau_{\gamma_2} = \Delta \tau_{\gamma'_2} + \Delta \tau$. Thereby, we select the events that occurs with a difference of $\Delta \tau$. Thus, the normalized state that describes these events is given by
\begin{eqnarray}
\vert{\Psi}\rangle &=& \frac{\sqrt{2}}{4} \int  \int d\omega_1 d\omega_2 f(\omega_1, \omega_2) \nonumber \\
&&  \left[\left( a_{1,1}^{\dagger}(\omega_1) + a_{1,2}^{\dagger}(\omega_1) \right)\left( a_{2,1}^{\dagger}(\omega_2) + a_{2,2}^{\dagger}(\omega_2) \right) \right. \nonumber \\
&& \left. e^{-i(\omega_{1}( \Delta \tau_{\gamma'_1}+ \Delta \tau ) + \omega_2 \Delta \tau_{\gamma'_1})} \right. \nonumber \\
&& \left.  +\left( a_{1,1}^{\dagger}(\omega_1) - a_{1,2}^{\dagger}(\omega_1) \right)\left( a_{2,1}^{\dagger}(\omega_2) - a_{2,2}^{\dagger}(\omega_2) \right) \right. \nonumber \\
&& \left. e^{-i(\omega_{1}(\Delta \tau_{\gamma'_2}+ \Delta \tau )+ \omega_2 \Delta \tau_{\gamma'_2})}e^{i(\alpha + \beta)} \right] \vert 0 \rangle_{1,2}.\nonumber \\
\end{eqnarray}

In this case, $\Delta \tau$ generates a global phase and consequently, the probability of detection becomes independent of $\Delta \tau$. It is interesting to note that the state obtained applying the conditions of indistinguishability can be understood as a state been seeing by two different clocks (they could be at different locations experiencing a different gravitational potential between them). Then, we could generate an entangled state even when the detectors are placed at different gravitational potentials.

To calculate the probability of detection  we define four projectors
\begin{eqnarray}
\widehat{P}_{i,\alpha} &=& \int d\omega a_{i,\alpha}^{\dagger}(\omega)\vert{0}\rangle\langle{0}\vert a_{i,\alpha}(\omega),\label{detectorC} 
\end{eqnarray}
where $\alpha$ denotes the output port at each Mach-Zehnder interferometer, and $\alpha=1,2$ denotes the Mach-Zehnder interferometer at left (1) or at right (2), respectively. The probability to detect a photon at output port $i$ at the Mach-Zehnder interferometer $\alpha=1$ and a photon at output port $j$ at the Mach-Zehnder interferometer $\alpha=2$ is given by $p_{i,j} = \rm{Tr}[|\Psi\rangle \langle \Psi |\widehat{P}_{i,1} \otimes \widehat{P}_{j, 2}]= \langle \Psi \vert \widehat{P}_{i,1} \otimes \widehat{P}_{j,2} \vert \Psi \rangle$. Thus, we have
\begin{eqnarray}
p_{i,j} &=& \frac{1}{8} \int d\omega'_1 d\omega'_2 d\omega_1 d\omega_2 f(\omega'_1,\omega'_2)^{\ast} f(\omega_1,\omega_2) \nonumber \\
&& \delta\left(\omega_1 - \omega'_1 \right) \delta\left(\omega_2 - \omega'_2 \right) \nonumber \\
&& \left[ e^{i \left( \omega_1 - \omega'_1 \right) (\Delta \tau_{\gamma'_1} + \Delta \tau) + i \left( \omega_2 - \omega'_2 \right) \Delta \tau_{\gamma'_1}} \right. \nonumber \\
&& \left. -(-1)^{\delta_{\alpha \alpha'}}  e^{i\left(\alpha + \beta \right)}  \right. \nonumber \\
&& \left.  e^{i\left( \omega_1 (\Delta \tau_{\gamma'_1} + \Delta \tau) - \omega'_1 (\Delta \tau_{\gamma'_2} + \Delta \tau) + \omega_2 \Delta\tau_{\gamma'_1} - \omega'_2 \Delta\tau_{\gamma'_2}\right)} \right. \nonumber \\
&&-(-1)^{\delta_{\alpha \alpha'}} \left. e^{-i\left( \alpha + \beta \right)} \right. \nonumber \\
&& \left. e^{-i\left( \omega_1 (\Delta \tau_{\gamma'_1} + \Delta \tau) - \omega'_1 (\Delta \tau_{\gamma'_2} + \Delta \tau) + \omega_2 \Delta \tau_{\gamma'_1} - \omega'_2 \Delta\tau_{\gamma'_2}\right)}  \right. \nonumber \\
&& \left. + e^{i \left( \omega_1 - \omega'_1 \right) (\Delta \tau_{\gamma'_2} + \Delta \tau) + i \left( \omega_2 - \omega'_2 \right) \Delta \tau_{\gamma'_2}} \right] \nonumber \\
&=& \frac{1}{4}\left( 1  -(-1)^{\delta_{ij}} \int d\omega_1 d\omega_2 \vert f(\omega_1,\omega_2) \vert^2  \right. \nonumber \\
&& \left. \cos \left( \omega_1 \left( \Delta \tau_{\gamma_1} - \Delta \tau_{\gamma_2}\right)+ \omega_2 \left( \Delta \tau_{\gamma'_1} - \Delta \tau_{\gamma'_2}\right) \right. \right. \nonumber \\
&& \left. \left. + \left( \alpha + \beta \right) \right) \right). \nonumber \\ \label{ProbDetFr2}
\end{eqnarray}
Defining  $\Delta \tau_{\gamma_2,\gamma_1} = \Delta \tau_{\gamma_2} -\Delta \tau_{\gamma_1}$ as the difference of proper time between paths $\gamma_1$ and $\gamma'_1$, the probability of detection Eq.\thinspace(\ref{ProbDetFr2}) becomes
Eq.\thinspace(\ref{ProbFransonUniversal}).

\end{document}